\documentclass[%
 reprint,
 superscriptaddress,
 amsmath,amssymb,
 aps,
 pre,
 floatfix,
]{revtex4-1}

\usepackage{graphicx} 
\usepackage{braket} 
\usepackage{extpfeil}
\usepackage{footnote}
\usepackage{footmisc}
\usepackage{xcolor}   

\usepackage{lipsum} 
\usepackage{dcolumn}  
\usepackage{bm}       
\usepackage{hyperref} 
\hypersetup{          
 pdftitle={},
 pdfauthor={Ohad Shpielberg},
 pdfsubject={},
 pdfkeywords={},
 unicode=true,
 breaklinks=true,
 linkcolor=blue, 
 citecolor=[rgb]{0.0,0.7,0.1},
 pdfborder={0 0 0},
 pdfborderstyle={},
 colorlinks=true,
 bookmarksnumbered=true,
 bookmarksopen=true,
 bookmarksopenlevel=1,
}

\usepackage{nicefrac,xfrac}

\usepackage[all]{xy}                        
\SelectTips{cm}{10}                         
\usepackage[abbreviations,xlatin]{foreign}  
\defasforeign[aesthetic]{\ae{}sthetic}      

\usepackage[normalem]{ulem}                 
\usepackage{comment}                        
\DeclareMathOperator{\Tr}{Tr}

\begin{document}


\title{Diffusion and entanglement in open  quantum systems}
%

\author{Ohad Shpielberg}
\email{ohad.shpilberg@college-de-france.fr}
\email{ohad19@gmail.com}
\affiliation{Coll\`{e}ge de France, 11 place Marcelin Berthelot, 75231 Paris Cedex 05 - France.}
\affiliation{Laboratoire de Physique Th\'{e}orique de l'\'{E}cole Normale Sup\'{e}rieure de Paris, CNRS, ENS  \& PSL Research University,UPMC  \& Sorbonne Universit\'{e}s, 75005 Paris, France.}

%

\date{\today} 


\begin{abstract}
The macroscopic fluctuation theory provides a complete hydrodynamic description  of nonequilibrium  classical diffusive systems. As a first step towards a diffusive theory of open quantum systems, we propose a microscopic open quantum system model -- the selective dephasing model. It exhibits genuine quantum diffusive scaling. Namely, the dynamics is diffusive and the density matrix remains entangled at large length scales and long time scales. 
\end{abstract}

\maketitle

\section{Introduction} 

Recently, transport properties of nonequilibrium quantum systems have been extensively studied  \cite{Eisert_Gogolin,Doyonintegrability,BertiniProfileXXZ}. Of particular interest are dissipative quantum systems \cite{ProsenXXZ,Ziolkowska_YB_Lindblad,MBLwBath_Nandkishore,MBL_Levi,Fischer2016,Medvedyeva2016,Firstenberg2020,QSSEP_Monthus,ZnidaricSpinDiffusionXXZ}, which in many cases lead to diffusive transport -- ubiquitous in both quantum  and classical physics.

The macroscopic fluctuation theory (MFT) \cite{MFT15}, a universal coarse grained theory, has been rigorously shown to completely encapsulate the dynamics of classical diffusive systems \cite{Bertini2002,Appert-Rolland2008,Shpielberg2018b}. The MFT allows  to  calculate nonequilibrium long range correlations \cite{BertiniCorrelations}, large deviations \cite{Additivity}, fluctuation induced forces \cite{Casimir}, predict dynamical phase transitions  \cite{BertiniLDF,AP_condition,BaekPRL,Baek16Jstat,OS_numerical}, their associated critical exponents \cite{Shpielberg2018b,Baek18PhysicaA,ABCtransitions}  and many more \cite{SadhuTracer,Agranov19}.
The present study is motivated by a salient question: Does a universal description of quantum diffusive dynamics, comparable to the MFT, exist?

Let us briefly recap the MFT. Rather than describing a diffusive  process at the mesoscopic scale through a master equation, the MFT takes a coarse grained approach where, away from criticality, only conserved fields survive. To make this more concrete,  consider a 1D diffusive system of interacting particles with particle density $n$. Conservation of particles in the bulk implies the continuity equation $\partial_t n = -\partial_x j$. Then, the diffusion equation results from the combination of the continuity equation and Fick’s law of the mean current  $j = -D \partial_x n + \sigma E$ where $D,\sigma$ are the diffusivity and conductivity encapsulating the dynamics and $E$ is an external field. To infer what are the correlations in the system, one must go beyond the mean description and  include fluctuations.  The MFT asserts that the fluctuations of the process at the coarse grained level are entirely described by supplementing the current in Fick’s law with a white noise term of magnitude proportional to the conductivity. Therefore, regardless of the microscopic description, the dynamics of the coarse grained mean behaviour as well as the fluctuations are entirely encapsulated by $D,\sigma$. This is the universal structure of the MFT.

A universal coarse grained theory of quantum diffusion is required to go beyond the classical MFT, e.g. to exhibit entanglement. First, it is necessary to provide a microscopic model that exhibits  entanglement and diffusive dynamics at the coarse grained level \cite{GullensShortEntanglement}, expanding the MFT's mean evolution structure. Then, one can verify whether the structure of the fluctuations remains universal.

 In this letter we introduce a microscopic model described by a Quantum master equation that demonstrates at the coarse grained level both diffusive dynamics and  entanglement at the steady state. The mean evolution is a set of non-local diffusion equations. This example demonstrates that a general theory describing diffusive open quantum system goes beyond the MFT. Furthermore, it provides us with a working example that will allow in turn to develop such a diffusive theory of open quantum systems.



\section{Setup}   Since simple diffusive systems are Markovian and dissipative, a convenient initial starting point would be to consider Markovian open quantum systems, given by the Gorini-Kossakowski-Sudarshan-Lindblad (GKSL) equation \cite{BreuerBook}. 
Consider a quantum system, coupled to a large environment with fast  relaxation times. The back-action of the system on the environment can thus be neglected on the relevant time scale of the system. The evolution of the system's density matrix follows   $\partial_t \rho = \mathcal{L}(\rho)$, where
\begin{eqnarray}
\label{eq:lindblad original }
\mathcal{L}(\rho) &=& - i \left[H,\rho\right] +  \sum_k L_{M_k} (\rho), 
\\ \nonumber
 L_M   (\rho)  &=& M \rho M^\dagger   - \frac{1}{2} \lbrace M^\dagger  M , \rho  \rbrace. 
\end{eqnarray}
Here $\left[ \cdot, \cdot  \right] $ and $\lbrace \cdot, \cdot \rbrace$ are correspondingly the commutation and anti-commutation relations. $H$ is a Hermitian operator and may contain both the system's Hamiltonian and its interaction with the environment.  The set of $M_k$ operators, orthonormal under the trace norm $ \Tr M^\dagger _k M_k' =\delta_{k,k'}$, 
are related to the interaction with the environments. 
One can also follow the evolution of operators in the Heisenberg picture, given by the adjoint equation $\partial_t O = \mathcal{L}^* (O)$.
If all the $M_k$ are Hermitian, $\mathcal{L}^* (\bullet) = +i [H,\bullet] + \sum_k L_{M_k}(\bullet)$.

To explore quantum diffusion, we introduce the ``selective dephasing'' model. Consider a periodic spin chain of $\Omega>2$ sites, with the Pauli matrices $\sigma^{x,y,z,\pm} _k$ operating at site $k$.  The selective dephasing model has the GKSL form \eqref{eq:lindblad original }. We set $H$  to be the XX Hamiltonian  
 $H_{\textnormal{XX}} = \varepsilon \sum^\Omega _{k=1 } \sigma^x _k \sigma^x _{k+1} + \sigma^y _k \sigma^y _{k+1}$ . The  operators $M_k =  \sqrt{\eta \nu_f} \sigma^z _k $  correspond to dephasing on the sites $k\in \mathcal{A}$, where $ \nu_f$ has units of inverse time and $\eta$ is a dimensionless parameter that controls the strength of the dephasing. Three choices of  $\mathcal{A}$, which select the dephasing sites, are considered: 
 \begin{gather}
  \mathcal{A}_\Omega \equiv  \lbrace 1,2,...,\Omega \rbrace,
   \quad 
   \mathcal{A}_{\lbrace \xi \rbrace} \equiv \lbrace j\in \mathcal{A}_\Omega | j\neq \xi,\xi+2 \rbrace, 
  \nonumber  \\ 
  \mathcal{A}_{\rm odd} \equiv \lbrace 1,3,5,...,\Omega-1 \rbrace,
 \end{gather}
 where for $\mathcal{A}_{\rm odd} $, $\Omega$ is assumed to be even. Note that the exact value of $\xi$ is irrelevant due to the periodic boundary conditions we consider. 
  This GKSL model, 
  can be derived  using the singular coupling limit for spins coupled to a set of  
  thermalized
   Bosonic baths \cite{BreuerBook} 
   and for Fermions in an optical lattice \cite{Sarker2012}.       

 The $ \mathcal{A}_{\Omega}$ model
has been exhaustively studied  \cite{Carollo2017,Eisler2011,Fischer2016,Bernard2018} and it even admits a Bethe ansatz solution \cite{Medvedyeva2016}. First, we recap the derivation of the diffusive dynamics of the $\mathcal{A}_\Omega$ model. 
The dephasing leaves no quantum fingerprints at long times. Then, we show that protecting some of the sites from dephasing as in the $\mathcal{A}_{\lbrace \xi \rbrace },\mathcal{A}_{\rm odd }$ models may maintain the diffusive dynamics and exhibits quantum fingerprints at long times. 

For the analysis, we define the operators  \begin{gather}
\label{eq:operator definitions}
P^\pm _k = \sigma^\pm _k \sigma^\mp _k ,   \quad 
\quad 
\mathbb{P}_k = P^+ _k  \prod_{l\in \mathcal{A}_\Omega/\lbrace k \rbrace } P^- _l  ,
\\ \nonumber 
\mathbb{S}^m _k =  (\sigma^+ _k \sigma^- _{k+m}
+\sigma^- _k \sigma^+ _{k+m} )
\prod_{l\in \mathcal{A}_\Omega/\lbrace k, k+m \rbrace }  P^- _l .
\end{gather}
 In the following, the long time dynamics of the selective dephasing models will be shown to be governed by these operators. Furthermore, the $\mathbb{S}^m _k$ will be used as witnesses of entanglement. 
 

\section{   $\mathcal{A}_\Omega$ dephasing  }  The operator evolution for $P^+ _i $ is given by the discrete conservation equation $\partial_t P^+ _i  = -J_i + J_{i-1}$, where $J_i  = 2i \varepsilon ( \sigma^+ _i\sigma^- _{i+1} - \sigma^- _i\sigma^+ _{i+1}   ) $. Therefore, on a ring  $ Q = \sum_{k\in \mathcal{A}_\Omega} \Tr P^+ _k \rho(s)  $ is conserved for any $\eta$. This is true for any $\mathcal{A}$ dephasing model.   
Trying to evaluate   $\partial_t J_i$ reveals that the equations do not close \cite{Bernard2018},  so this naive approach fails. Fortunately, the dynamics simplifies as states not in the kernel of   $\sum_{k \in \mathcal{A}_\Omega} L_{M_k} $ are quickly filtered \cite{Fischer2016,Bernard2018,suppmat}. States in the kernel are called pointer states. The resulting long time dynamics, quickly attained for large $\eta$ values, is governed by transitions between these pointer states. Hence, we derive a perturbative expression for the pointer states dynamics at large $\eta$ for a general GKSL equation of the form 
\begin{eqnarray}
\label{eq:strong Lb form}
\mathcal{L}(\rho) &=& \mathcal{L}_S (\rho) + \eta \mathcal{L}_b (\rho),
\\ \nonumber 
\mathcal{L}_S (\rho) = - i [H,\rho ], & &  \mathcal{L}_b (\rho) = \nu_f \sum_{k\in \mathcal{A}} L_{M_k}(\rho).
\end{eqnarray}
For large $\eta$, the density matrix quickly converges into a mixture of  pointer states -- the states  in the kernel of $\mathcal{L}_b$. Then, the density matrix can only evolves on a long time scale \cite{Cai2013,Marcuzzi2014,Degenfeld-Schonburg2014}. A slow dynamics, where the density matrix spreads to other pointer states   emerges if $ \Pi_0 \mathcal{L}_S \Pi_0 =0$ and  $\mathcal{L}_b = \sum_{\nu\leq0} \nu \Pi_\nu$. Here $\Pi_\nu$  are projectors and specifically $\Pi_0$ is the projector into the kernel. To leading order $\partial_s\rho=\mathfrak{u}    \rho 
$, where   \cite{suppmat,Fischer2016,Bernard2018}
\begin{eqnarray}
\label{eq:effective evo}
\mathfrak{u}     &=& - \Pi_0  \mathcal{L}_S (\mathcal{L}^\perp _b)^{-1} \mathcal{L}_S , 
\end{eqnarray}
with the rescaled time $s=t/\eta$ at the limit of  $t,\eta \rightarrow \infty$. 
$(\mathcal{L}^\perp _b)^{-1} = \sum_{\nu<0} \nu^{-1} \Pi_\nu$ is the inverse to the restriction of $\mathcal{L}_b$ outside of the kernel. 
The above assumptions are valid in all the  variants of the selective dephasing model \cite{suppmat}. Finding the projector to the kernel $\Pi_0$ may be easier than finding a basis for the pointer states as one needs to worry about positivity and unit trace.  For Hermitian $M_k$, the operator evolution is  $\partial_s O = \mathfrak{u} O$. Hence, dealing with pointer operators, i.e. $O = \Pi_0 O$ may be easier than dealing with pointer states.  

To analyze the effective dynamics, it is essential to identify the kernel of $\mathcal{L}_b$. For $\mathcal{A}_\Omega$, this kernel is spanned by the $2^\Omega$ pointer states $\prod_k P^{\pm} _k$. One can  give a classical  interpretation to the effective dynamics of  \eqref{eq:effective evo} in this case. Interpreting the operators $P^\pm$ as  projectors into an occupied (empty) lattice site state implies that the pointer states span all the configurations of a lattice gas. Namely, each site is either occupied by a particle or not. 
The dynamics corresponds to the simple symmetric exclusion process (SSEP), where particles can jump to empty neighboring lattice sites with rate $D= \nicefrac{2\varepsilon^2}{\nu_f}$ \cite{Derrida2007,Mallick2015}. The SSEP is diffusive and classical. 
So, we have gained no  insight on the behaviour of diffusive open quantum systems. 

By protecting  some of the sites from the dephasing noise, one expects a larger kernel, which in turn could lead to richer dynamics. This is the reasoning behind the   $\mathcal{A}_{\lbrace \xi \rbrace} , \mathcal{A}_{\rm odd}$ models. We restrict the study here to the already interesting case $Q=1$. For the analysis of the qualitatively similar $Q\neq1$ see \cite{suppmat}.

\section{$\mathcal{A}_{\lbrace \xi \rbrace}$  dephasing}
 The kernel becomes larger than in the $\mathcal{A}_\Omega$ case as it contains, e.g. the pointer operator
$\mathbb{S}^2 _\xi  $ (see \eqref{eq:operator definitions}), as well as the states $\mathbb{P}_k$. 
The effective dynamics for the $\mathcal{A}_{\lbrace \xi \rbrace}$ dephasing provides us with a closed set of pointer operator equations 
\begin{eqnarray}
\label{eq:selective dephasing 2 EoM}
\partial_s \mathbb{P}_k &=& D \Delta \mathbb{P}_k, \quad k \neq \xi-1,\xi,\xi+1,\xi+2
\nonumber \\\nonumber
\partial_s \mathbb{P}_k &=& 2D \Delta \mathbb{P}_k + D \mathbb{S}^2 _\xi, \quad k = \xi,\xi+2  
\\\nonumber
\partial_s \mathbb{P}_{\xi+1} &=& 2D \Delta \mathbb{P}_{\xi+1} - 2D\mathbb{S}^2 _{\xi}    
\\
\partial_s \mathbb{P}_{\xi-1} &=& D \Delta \mathbb{P}_{\xi-1} +D(\mathbb{P}_{\xi}-\mathbb{P}_{\xi-1}) 
\\ \nonumber
\partial_s \mathbb{S}^2 _\xi &=& -2D \Delta \mathbb{P}_{\xi+1} -4D\mathbb{S}^2 _\xi,
\end{eqnarray}
where $\Delta$ is the discrete Laplacian operator. 
  
  Note that in the slow dynamics regime, operators outside of the kernel have already decayed in the Heisenberg picture, i.e. $\partial_s O =0$ if $\Pi_0 O =0$. Therefore, more information on the slow dynamics can be recovered from the evolution equations of the (non-pointer operators) $\mathbb{S}^2 _{k\neq\xi}$. Since  $ \Pi_0\mathbb{S}^2 _{k\neq\xi} =0 $, we obtain a set of  identities  
\begin{eqnarray}
\label{eq:Axi identities}
\Delta \mathbb{P}_{k+1}&=&0, \quad |k-\xi|\geq 3 
\nonumber \\
\Delta \mathbb{P}_{\xi+3} &=& \mathbb{P}_{\xi+3} - \mathbb{P}_{\xi+2} 
\nonumber \\
\Delta \mathbb{P}_{\xi-1} &=& \mathbb{P}_{\xi-1} - \mathbb{P}_{\xi} 
\nonumber \\  
\mathbb{S}^2 _\xi &=& \Delta \mathbb{P}_\xi = \Delta\mathbb{P}_{\xi+2}. 
\end{eqnarray}
The slow dynamics identities \eqref{eq:Axi identities} are at the operator level and hence do not depend on initial conditions. 
We denote by $S_{\xi}(s) = \Tr \,  \mathbb{S}^2 _\xi \rho(s) $ and
$q_k (s) = \Tr \mathbb{P}_k \rho(s) $  the expectation values associated with the pointer states. From the $\Omega-1$ identities \eqref{eq:Axi identities} and from the conserved $Q=1= \sum_k q_k $, we recover a single solution $q_{k\neq \xi+1} = \frac{1}{\Omega} (1-S_\xi) $ and $q_{\xi+1} = \frac{1}{\Omega}( 1 + (\Omega-1)S_\xi)$. 
The density matrix of the effective dynamics is a (positive semidefinite, Hermitian and trace 1) combination of the operators $\mathbb{P}_k$ and $ \mathbb{S}^2 _\xi$. The Peres-Horodecki criterion asserts that a two two-level system is entangled if and only if the partial transpose to the density matrix is not positive definite  \cite{Peres1996,suppmat}.  We use the Peres-Horodecki criterion to probe for bipartite entanglement in our system. In the effective dynamics bipartite entanglement can occur in the $\mathcal{A}_{\lbrace \xi \rbrace }$ model only between the spins in $\xi,\xi+2$ and only if  $S_\xi \neq 0$  \cite{suppmat}. Hence, $S_\xi$ serves as a witness of entanglement in the system. 


 We evaluate the   values $S_\xi$ and $q_k$ using the evolution equations \eqref{eq:selective dephasing 2 EoM}. They corresponds to the trivial solution $S_\xi=0$ and $q_k =1/\Omega$ in the effective dynamics. Hence,  entanglement quickly decays.  This suggests, as expected, that a microscopic protection from the environment is not enough to retain an interesting hydrodynamic quantum behaviour (see \cite{suppmat} for numerical verification).

\section{$\mathcal{A}_{\rm odd}$  dephasing} 
\begin{figure}
\begin{center}
 \includegraphics[width=0.35\textwidth]{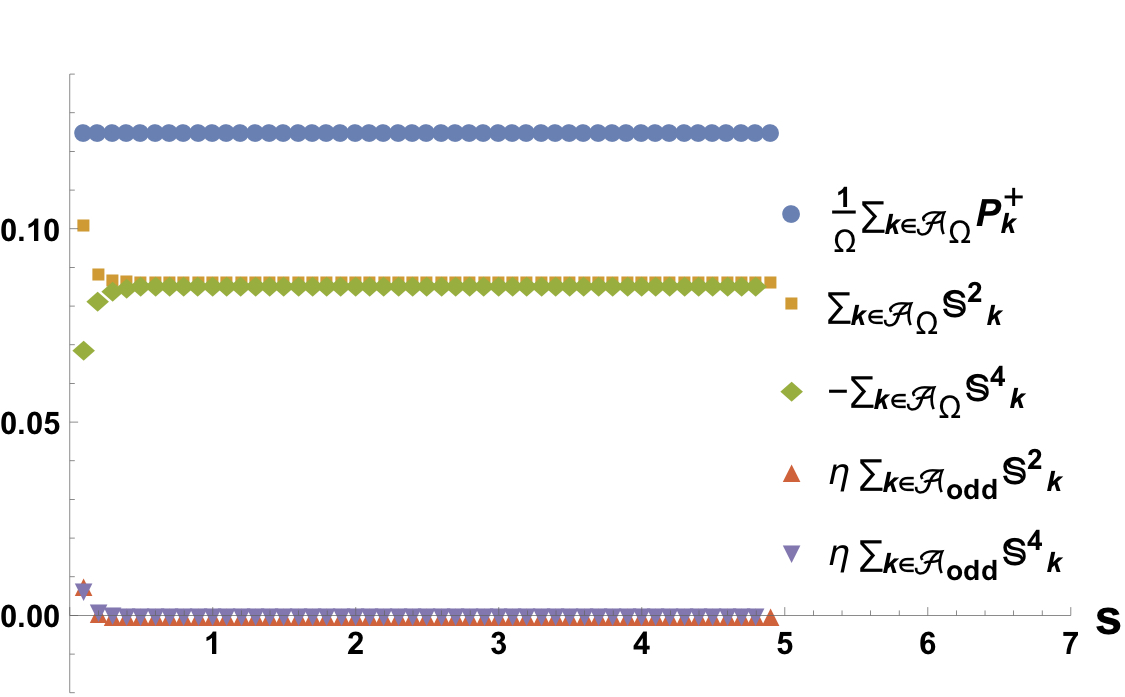}
\includegraphics[width=0.35\textwidth]{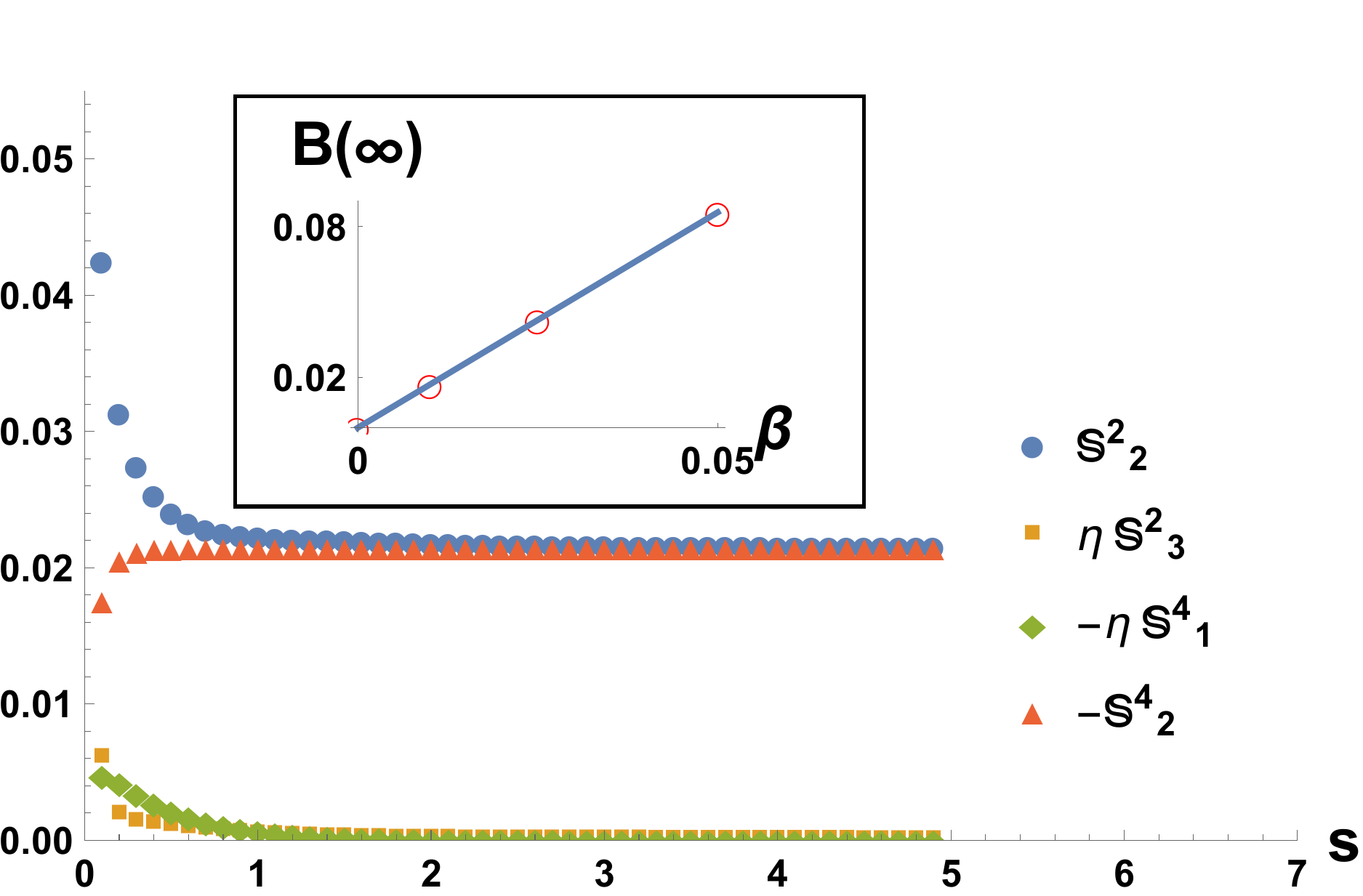}

\caption{\label{fig:S2 quasi conserved} 
The numerical evaluation directly evolves the density matrix for the $\mathcal{A}_{\rm odd}$ selective dephasing model with $\Omega=8$ sites, $\eta = 10, \varepsilon=\nu_f=1$. The initial density matrix is $\rho_0 = \frac{1}{\Omega}\sum_{k\in \mathcal{A}_\Omega} \mathbb{P}_k + \beta ( \mathbb{S}^2 _2 +2\mathbb{S}^2 _4 - \mathbb{S}^2 _6 + \mathbb{P}_2- \frac{2}{3} \mathbb{P}_3  -\frac{1}{3}\mathbb{P}_7  )$. \textbf{Top}: The expectation values of the conserved $\sum_{k \in \mathcal{A}_\Omega}P^+ _k$,   effectively conserved $\sum_{k\in \mathcal{A}_\Omega}\mathbb{S}^2 _k, \sum_{k\in \mathcal{A}_\Omega}\mathbb{S}^4 _k$ and decaying quantities $\sum_{k\in \mathcal{A}_{\rm odd}}\mathbb{S}^2 _k,\sum_{k\in \mathcal{A}_{\rm odd}}\mathbb{S}^4 _k$. The decaying quantities are multiplied by a factor $\eta$ to highlight their  quick decay. \textbf{Bottom}: The expectation values associated with long lasting bipartite entanglement. Operators not in the kernel are multiplied by a factor $\eta$ to highlight their quick decay.  \textbf{Inset}: The steady state value  $B(t) = \sum_{k\in \mathcal{A}_{\Omega}} \Tr \rho(t) \mathbb{S}^{2} _k $ as a function of $\beta$ in the initial conditions. Starting at $B(0)=4 \beta$, the plot shows $B(\infty)=1.71 \beta$ which demonstrates the lasting entanglement in the system. 
}
\end{center}
\end{figure}

As before, we study the effective dynamics at large $\eta$ to obtain the evolution within the kernel. Notice that $\mathbb{P}_k $ for $k\in \mathcal{A}_\Omega$ and the operators $\mathbb{S}^{2m} _{2k}$ are  in the kernel for  $k,m=1,2,...,\Omega/2$. For  $Q=1$, we obtain a set of evolution equations and identities corresponding to  the operators in the kernel. Here, we keep track also of  $ \mathbb{S}^{2m} _k $ for odd $k$, even though they are not in the kernel and thus have vanishing expectation values in the effective dynamics. This highlights the emerging diffusive picture,  but does not change any expectation value (see Fig.\ref{fig:S2 quasi conserved} and \cite{suppmat}).  The (diffusive) evolution equations are
\begin{eqnarray}
\label{eq:evolution Aodd}
\partial_s \mathbb{P}_{k} &=&  2D \Delta  \mathbb{P}_{k}  -  D \Delta \mathbb{S}^2 _{k-1}
\\ \nonumber 
\partial_s \mathbb{S}^2 _{k} &=&  -2D\Delta  \mathbb{P}_{k+1} + 2D\Delta \mathbb{S}^2 _{k}-D\Delta \mathbb{S}^4 _{k-1}
\\ \nonumber 
\partial_s \mathbb{S}^{2+2m} _{k} &=&   - D\Delta \mathbb{S}^{2m} _{k+1}
+2D\Delta \mathbb{S}^{2m+2} _{k}- D\Delta \mathbb{S}^{4+2m} _{k-1},
\end{eqnarray}
for $2\leq 2+2m \leq \Omega/2$. The periodicity of the system helps to truncate the equations \eqref{eq:evolution Aodd} as $\mathbb{S}^{\Omega/2 +2 } _k = \mathbb{S}^{\Omega/2 - 2} _{k+ (\Omega/2 +2)} $ for $\Omega/2$ even, and   $\mathbb{S}^{\Omega/2 +1 } _k = \mathbb{S}^{\Omega/2 - 1} _{k+ (\Omega/2 +1)} $ for $\Omega/2$ odd. 

To find the steady state solution, we use the translational invariance of the system (in discrete  jumps of 2). Define $q_{\rm o} = \Tr \rho(s) \mathbb{P}_{2k+1}$, $q_{\rm e} = \Tr \rho(s) \mathbb{P}_{2k}$ and $S^{2m} =\Tr \rho(s) \mathbb{S}^{2m} _{2k} $. From \eqref{eq:evolution Aodd} and for $\Omega/2$ even, we find the steady state values $S^{2m} = -S^{2m+2} $ as well as $ q_{\rm o}-q_{\rm e} = \frac{1}{2} S^2  $. This is numerically confirmed in \cite{suppmat}.

Similarly to the $A_{\lbrace\xi\rbrace}$ dephasing, the expectation value of $S^{2m}$ is a witness of entanglement in the steady state.  Note that the steady state density matrix is constructed using the pointer operators $\mathbb{P}_k , \mathbb{S}^{2m} _{2k}$. Then, using the Peres-Horodecki criterion shows that any two even sites $2k',2k''$ are bipartite entangled if the expectation of $\mathbb{S}^{2(k''-k')} _{2k'}$ is non-vanishing \cite{suppmat}.


For $\Omega/2$ odd, periodicity implies that  $S^{\Omega/2+1} = S^{\Omega/2-1}  $. Then, the steady state equations result in  $S^{2m}=0, q_{\rm o}=q_{\rm e} = 1/\Omega$ and thus the system will not be entangled at the steady state.  In what follows, we only study  $\Omega/2$ even, where the entanglement can survive at the steady state. The effective evolution equations \eqref{eq:evolution Aodd}  suggests that the expectation values of $\sum_{k \in \mathcal{A}_\Omega}\mathbb{S}^{2m} _{k} $ are conserved quantities in the effective dynamics. However, this is not the case for the $t$ time evolution. It implies that only beyond a transient period $t \gtrsim \eta \nu_f/ \varepsilon^2$,    $\sum_{k\in \mathcal{A}_\Omega}\mathbb{S}^{2m} _{k}$ is conserved.

Using the effective conservation, we define a witness of entanglement in the system $B(t) = \sum_{k\in\mathcal{A}_\Omega}\Tr \rho(t) \mathbb{S}^2 _k $. Using this witness, Fig.~\ref{fig:S2 quasi conserved} and \cite{suppmat}  numerically demonstrates that entanglement survives the long time limit for generic initial conditions,  provided the initial conditions have some nonzero entanglement value (other initial conditions were tested with similar results). Notice that the entanglement survives on all length scales of the system. That is, the bipartite entanglement spreads to all length scales even if it is initially restricted to a local region.  Hence, we have successfully found a model where long range entanglement can be observed for diffusive transport.


Contrary to the classical diffusion equation description, the hydrodynamic equations for the quantum diffusion of the $\mathcal{A}_{\rm odd}$ dephasing model can be expected to be non-local due to the long ranged entanglement. 
For $\Omega/2$ even,  we  define the coarse grained length $x=j/\Omega$, and time   $\tau = s D / \Omega^2 $ scales as well as the coarse grained operators $\mathbb{P}(x,t),\mathbb{S}^{2m}(x,\tau)$.  Using the vector representation  $v = \left( \mathbb{P}, \mathbb{S}^2 , \mathbb{S}^4 , ... \mathbb{S}^{\Omega/2} \right)$, 
  the non-local hydrodynamic diffusion equation stemming from \eqref{eq:evolution Aodd}  is
\begin{eqnarray}
\label{eq:Diffusion matrix}
    \partial_\tau v(x) = \partial_{xx} \mathcal{D}_0 v(x) + \partial_{xx} \mathcal{D}_1 v(x+\Omega/2),
\end{eqnarray}
where 
\begin{eqnarray*}
\mathcal{D}_1 &=& \begin{pmatrix}
    0  & ... &   &   
    \\ 
    \vdots  & \ddots &   &
    \\
      &  & 0  &0   
    \\
    &  & -1  &0   
    \end{pmatrix},
    \end{eqnarray*}
    \begin{eqnarray*}
    \mathcal{D}_0 &=& \begin{pmatrix}
    2  & -1 & 0  & 0  & 0   & ... & 0
    \\ 
    -2 & 2  & -1  & 0  & 0   & ... & 0
    \\
    0  & -1 & 2  & -1 & 0   & ... & 0
    \\
    0 & 0   & -1 & 2  & -1  & ... & 0
    \\
    \vdots& & \ddots & \ddots& & &  
    \\
    0 & 0   & ... &   &  0 & -1 & 2
        \end{pmatrix}. 
\end{eqnarray*}
One can show, using Fourier analysis and the Perron-Frobenius theorem, that the system of equations is stable.

\section{Summary}  
 In this letter we have introduced and studied the dynamics of the selective dephasing model with three selective dephasing schemes.  Accentuating the dephasing part of the dynamics allows analytical treatment of the long time dynamics.

 
 A local protection from dephasing, as in the $\mathcal{A}_{\lbrace \xi \rbrace}$ dephasing model,  is not sufficient to keep quantum fingerprints at the steady state. The dephasing only at odd sites together with the Hamiltonian dynamics results in effectively conserved numbers. These new conserved numbers are witnesses of entanglement at the steady state at any length scale. Furthermore,   the coarse grained dynamics is encapsulated in \eqref{eq:Diffusion matrix}, a hydrodynamic non-local diffusion equation.



 Note that the effective dynamics cannot predict the steady state values of the conserved numbers from the initial state. This is due to the microscopic time $t\lesssim \eta \nu_f/\varepsilon^2$, where there is no conservation. However, we stress that the effectively conserved numbers generally do not vanish if some bipartite entanglement is introduced in the initial condition    (Fig.\ref{fig:S2 quasi conserved} and \cite{suppmat}). 
We further note that the perturbation theory in 
\eqref{eq:effective evo} may break the  quantum dynamical semigroup structure \cite{Alicki}. For $Q=1$, we verify that effective evolution of the density matrix in the pointer state subspace satisfies a GKSL equation \eqref{eq:lindblad original } according to $\partial_s \Pi_0 \rho  = \frac{1}{\nu_f} \Pi_0 L_{H_{XX}} (\rho) $. Further evidence for the validity of the perturbation theory is provided in \cite{suppmat}. 

The study of the effective dynamics of the selective dephasing model was restricted to its simplest case $Q=1$. Onerous treatment is required for  $Q\neq 1$, but similar results, i.e. diffusion and entanglement, are obtained \cite{suppmat}.

 It is now clear that a coarse grained theory of diffusive open quantum systems must allow entanglement on all scales resulting in non-local evolution equations.
 Thus, such a theory will differ from the classical MFT, which is considered a local theory. 
 
 The present study, in particular the odd dephasing model, sets the stage for studying whether dynamics in diffusive quantum systems is indeed universal as its classical counterpart. Here, we considered the GKSL equation, which is an average equation. Following the degrees of freedom of the environment allows to inspect  quantum trajectories, e.g.  through quantum Langevin equations  \cite{KnightReview,ZollerGardiner}. Finding whether the coarse grained description of the quantum trajectories is universal will be an important step towards a nonequilibrium theory of diffusive open quantum systems.

Recall that the selective dephasing model can be derived from a Hamiltonian dynamics through the singular coupling limit. Therefore, it is of interest to implement the selective dephasing model $\mathcal{A}_{\rm odd}$ using cutting-edge experimental techniques \cite{Weimer_RydbergSimulator,Hild_NoneqHeisenberg,Boll_SpinDensityFermiHubbard,Helmrich_QmasterExp}. How to take advantage of the entanglement propagation to all scales of the selective dephasing model is yet another open challenge.

\begin{acknowledgments}
Denis Bernard,  Federico Carollo,  Yaroslav Don, Juan P. Garrahan, Rob Jack, Toni Jin, Igor Lesanovsky,  Katarzyna Macieszczak and Naftali Smith are acknowledged for fruitful discussions.  The support of grant ANR-14-CE25-0003 is  appreciated.

\end{acknowledgments}


\bibliographystyle{apsrev4-1}   
\bibliography{main.bib} 


\clearpage
\newpage

\setcounter{page}{1}
\setcounter{section}{0}

\onecolumngrid
\begin{center}
{\large \bf supplementary material for ``Diffusion and entanglement in open  quantum systems''}

\vspace{0.5cm}

Ohad Shpielberg

\vspace{0.5cm}

\end{center}

\onecolumngrid

\section{ Derivation of the perturbation theory  \label{sec:app:perturnation theory }}

The purpose of this section is to derive the effective dynamics in the large $\eta$ limit for the GKSL setup (4)
.  The density matrix is assumed to have the perturbative expansion 
\begin{equation}
\rho = \rho_0 + \frac{1}{\eta} \rho_1 + \frac{1}{\eta^2} \rho_2 + ... \label{eq:rho exp}
\end{equation}
Using \eqref{eq:rho exp} in 
(4), we find 
\begin{eqnarray}
0 = \mathcal{L}_b (\rho_0) \label{eq:order exp. 1}\\
\partial_t \rho_0 = \mathcal{L}_S(\rho_0) + \mathcal{L}_b (\rho_1) \label{eq:order exp. 0}\\ 
\partial_t \rho_1 = \mathcal{L}_S(\rho_1) + \mathcal{L}_b (\rho_2) 
\label{eq:order exp. -1}
\label{eq:order exp. -2}
\end{eqnarray}
We assume that the discrete decomposition  $\mathcal{L}_b = \sum_{\nu \leq 0} \nu \Pi_\nu$ exists, where $\Pi_\nu$ are projectors. Moreover, let us assume that $\Pi_0 \mathcal{L}_S \Pi_0 =0$, which facilitates the slow dynamics.

From \eqref{eq:order exp. 1}, we find that $\rho_0 \in \textnormal{Ker}\mathcal{L}_b$ so that $\rho_0 = \Pi_0 \rho_0$. Projecting \eqref{eq:order exp. 0} onto $\Pi_0$ and using our assumption leads to 
\begin{equation}
\partial_t \rho_0  = \Pi_0 \mathcal{L}_S \Pi_0 \rho_0 = 0.  
\label{eq:slow dynamics result}
\end{equation}
Next, we define $\rho^\parallel _k \equiv  \Pi_0 \rho_k$ and $\rho_k \equiv  \rho^\parallel _k +\rho^\perp _k $. Then, using \eqref{eq:slow dynamics result} and \eqref{eq:order exp. 0}, we find that $\rho^\perp _1 = - (\mathcal{L}^\perp _b)^{-1} \mathcal{L}_S  \rho_0 $. This suggests that  $\partial_t \rho^\perp _1 = 0$. Projecting \eqref{eq:order exp. -1} onto $\Pi_0$ and using the $\rho^\perp _1$ expression  gives
\begin{equation}
    \partial_t \rho^\parallel _1 =  - \Pi_0 \mathcal{L}_S (\mathcal{L}^\perp _b)^{-1} \mathcal{L}_S  \rho_0.
    \label{eq:rho1 par evo}
\end{equation}
Combining \eqref{eq:rho1 par evo} with $\partial_t \rho ^\perp _1 = 0 $ implies that 
\begin{eqnarray}
\partial_t \rho_1 &=&     \mathfrak{u} \rho_0.  
    \\ \nonumber
     \mathfrak{u} &=& - \Pi_0 \mathcal{L}_S (\mathcal{L}^\perp _b)^{-1} \mathcal{L}_S 
\end{eqnarray}
 So, in the rescaled time $s=t/\eta$, we find up to $1/\eta$ corrections the evolution equation
\begin{equation}
    \partial_s \rho = \mathfrak{u} \rho.
\end{equation}


\section{ Perturbation theory assumptions }

The previous section presented the perturbation theory for the strong $\eta$ limit. There, we assumed that $\Pi_0 \mathcal{L}_S\Pi_0 =0$ and that $\mathcal{L}_b = \sum_{\nu\leq0} \nu \Pi_\nu$. The purpose of this section is to directly show that the assumptions are valid for the selective dephasing model. 

First, consider a general density matrix $\rho$ for the spin chain with $\Omega$ sites. For any density matrix, we can always uniquely find the coefficients $r_{\vec{\alpha}}$ such that $\rho = \sum_{\vec{\alpha}} r_{\vec{\alpha}} A_{\vec{\alpha}}  $ and 
\begin{equation}
    A_{\vec{\alpha}} =
    A_{\alpha_1}\otimes A_{\alpha_2}\otimes ... \otimes A_{\alpha_\Omega},  
\end{equation}
with $\vec{\alpha} = (\alpha_1,...,\alpha_\Omega)$ and $A_{\alpha_i}= \lbrace \sigma^\pm, P^\pm \rbrace$ for $\alpha_i = {\pm s ,\pm p}$ correspondingly. It is straight-forward to show that  
\begin{equation}
\mathcal{L}_b A_{\vec{\alpha}} = -2 \nu_f \sum_{k \in \mathcal{A}} (\delta_{\alpha_k,+s} + \delta_{\alpha_k,-s})  A_{\vec{\alpha}}.  
\end{equation}
Namely, we can write $\mathcal{L}_b = \sum_{\nu\leq 0} \nu \Pi_\nu$ where for any $\vec{\alpha}$ there is a corresponding eigenvalue   $\nu(\vec{\alpha})=-2 \nu_f \sum_{k \in \mathcal{A}} (\delta_{\alpha_k,+s} + \delta_{\alpha_k,-s}) $ and the projectors $\Pi_{\nu'} A_{\vec{\alpha}} = \delta_{\nu', \nu({\vec{\alpha}})} A_{\vec{\alpha}}  $. For any choice of $\mathcal{A}$ as in the main text, the set of eigenvalues $\nu$ is non-positive and finite. Thus, we have justified the assumption  $\mathcal{L}_b = \sum_{\nu\leq 0} \nu \Pi_\nu$. 

Next, let us show that indeed $\Pi_0 \mathcal{L}_S \Pi_0 =0  $ for the selective dephasing. Since $\mathcal{L}_S(A_{\vec{\alpha}}) = -2\varepsilon i\sum_k \left[h_k ,A_{\vec{\alpha}} \right] $ with $ h_k = \sigma^+ _k  \sigma^- _{k+1} + \sigma^- _{k}  \sigma^+ _{k+1}    $, it is sufficient to show that 
$\Pi_0 \left[h_k , A_{\vec{\alpha}}  \right] =0  $ 
for any  $\vec{\alpha}$ such that $\nu(\vec{\alpha})=0$.

First, we show that for $\mathcal{A}_\Omega$, where the kernel is spanned by $ P^{\epsilon_1} \otimes P^{\epsilon_2} \otimes ...   $ with  $\epsilon_k = \pm1$. It is straightforward to verify that 
\begin{equation}
\label{eq:identity 1}
    \left[ h_k , 
  P^{\epsilon_k} _{k}  P^{\epsilon_{k+1}} _{k+1}  
    \right]  = \epsilon_{k}\epsilon_{k+1}(\sigma^+ _k  \sigma^- _{k+1} (\delta_{\epsilon_k,+}-\delta_{\epsilon_{k+1},+})
    + \sigma^- _{k}  \sigma^+ _{k+1}(\delta_{\epsilon_k,-}-\delta_{\epsilon_{k+1},-})).
\end{equation}
So, it is clear that for $\nu(\vec{\alpha})=0$, $\Pi_0\left[h_k,A_{\vec{\alpha}}\right]=0$. Therefore, for the selective dephasing with $\mathcal{A}_\Omega$, we find that indeed $\Pi_0\mathcal{L}_S\Pi_0 =0$. 

For the $A_{\xi}$ selective dephasing, $A_{\vec{\alpha}}$ is in the kernel if $\alpha_{k\neq \xi,\xi+2} = {\pm p } $ and   $\alpha_{k= \xi,\xi+2} $ can take any of the values $\lbrace \pm s, \pm p \rbrace $. Using the identities 
\begin{eqnarray}
\label{eq:identity 2}
\left[ h_k , 
  P^{\epsilon_k} _{k}  \sigma^{\epsilon_{k+1}} _{k+1}  
    \right]  &=& 
    -\epsilon_{k} \sigma^{\epsilon_k} _k ( \delta_{\epsilon_k,\epsilon_{k+1}}(P^+ _{k+1}-P^- _{k+1}) + \epsilon_{k+1}P^{-\epsilon_{k+1}} _{k+1}  )
    \\
    \left[ h_k , 
  \sigma^{\epsilon_k} _{k}  P^{\epsilon_{k+1}} _{k+1}  
    \right]  &=&
    \epsilon_{k+1} \sigma^{\epsilon_{k}} _{k+1} ( \delta_{-\epsilon_k,\epsilon_{k+1}}(P^+ _{k}-P^- _{k}) - \epsilon_{k}P^{\epsilon_{k}} _{k}  )
\end{eqnarray}
and \eqref{eq:identity 1} we indeed see that $\Pi_0 \mathcal{L}_S(A_{\vec{\alpha}})$ vanishes for $\nu(\vec{\alpha})=0$ as non-vanishing terms have $\sigma^\pm$ terms in odd-labeled sites. 

For the $A_{\rm odd}$ selective dephasing, $A_{\vec{\alpha}}$ is in the kernel if $\alpha_{k \in A_{\rm odd} } = {\pm p } $ and   $\alpha_{k \notin A_{\rm odd}} $ can take any of the values $\lbrace \pm s, \pm p \rbrace $. Using the identities \eqref{eq:identity 1} and \eqref{eq:identity 2},  and  similarly to $\mathcal{A}_{\lbrace \xi \rbrace}$, we see that $\Pi_0 \mathcal{L}_S(A_{\vec{\alpha}})$ vanishes for $\nu(\vec{\alpha})=0$ as again, non-vanishing terms have $\sigma^\pm$ terms in odd-labeled sites.

To conclude, we have verified that the assumptions for the perturbation theory are justified for the selective dephasing models  $\mathcal{A}_\Omega,\mathcal{A}_{\lbrace \xi \rbrace} ,\mathcal{A}_{\rm odd}$


\section{ Numerical support for the   $\mathcal{A_{\lbrace \xi \rbrace}} $ dephasing model    }

The purpose of this section is to numerically corroborate the predictions of the effective dynamics for the $\mathcal{A}_{\lbrace \xi \rbrace} $ dephasing model. 

The numerical protocol  directly evolves the  density matrix evolution according to equation (1) for the $\mathcal{A}_{\lbrace \xi \rbrace} $ selective dephasing model. We numerically evolve the density matrix  for $\Omega=8$ sites  with $\beta = 0.05, \epsilon=\nu_f=1,\eta=10$ with $\xi=2$ and starting at the initial density matrix
\begin{equation}
\rho_0 = \frac{1 - \beta/2}{\Omega} \sum_{k\in \mathcal{A}_\Omega} \mathbb{P}_k
+
 \frac{\beta}{2} (\mathbb{S}^2 _2  + \mathbb{P}_1  + 2 \mathbb{P}_2 - 3 \mathbb{P}_3 - 4 \mathbb{P}_4 + \frac{3}{2} \mathbb{P}_5  
     + 3 \mathbb{P}_6 + \frac{1}{2} \mathbb{P}_8).
     \end{equation}

In the main text we assert that for $Q=1$,  $q_k = 1/\Omega $  and $S_\xi = 0$ in the effective dynamics. This is corroborated in Fig.~\ref{fig:Conservation_Axi}.  


To show convergence to the effective dynamics at large $\eta$, we plot the errors of $q_k,S_\xi$ from the expected values for various $\eta$. For $E=q_k,S_\xi$ we define the relative error as  $|E(s)-\langle E \rangle |/|E(0)|$ where $\langle E \rangle = \Omega^{-1},0$ for $q_k,S_\xi$ correspondingly. In Fig.~\ref{fig:Error_est_Axi} the relative errors are shown to converge as $\eta$ increases.

\setcounter{figure}{2}    

\begin{figure}
\begin{center}
 \includegraphics[width=0.48\textwidth]{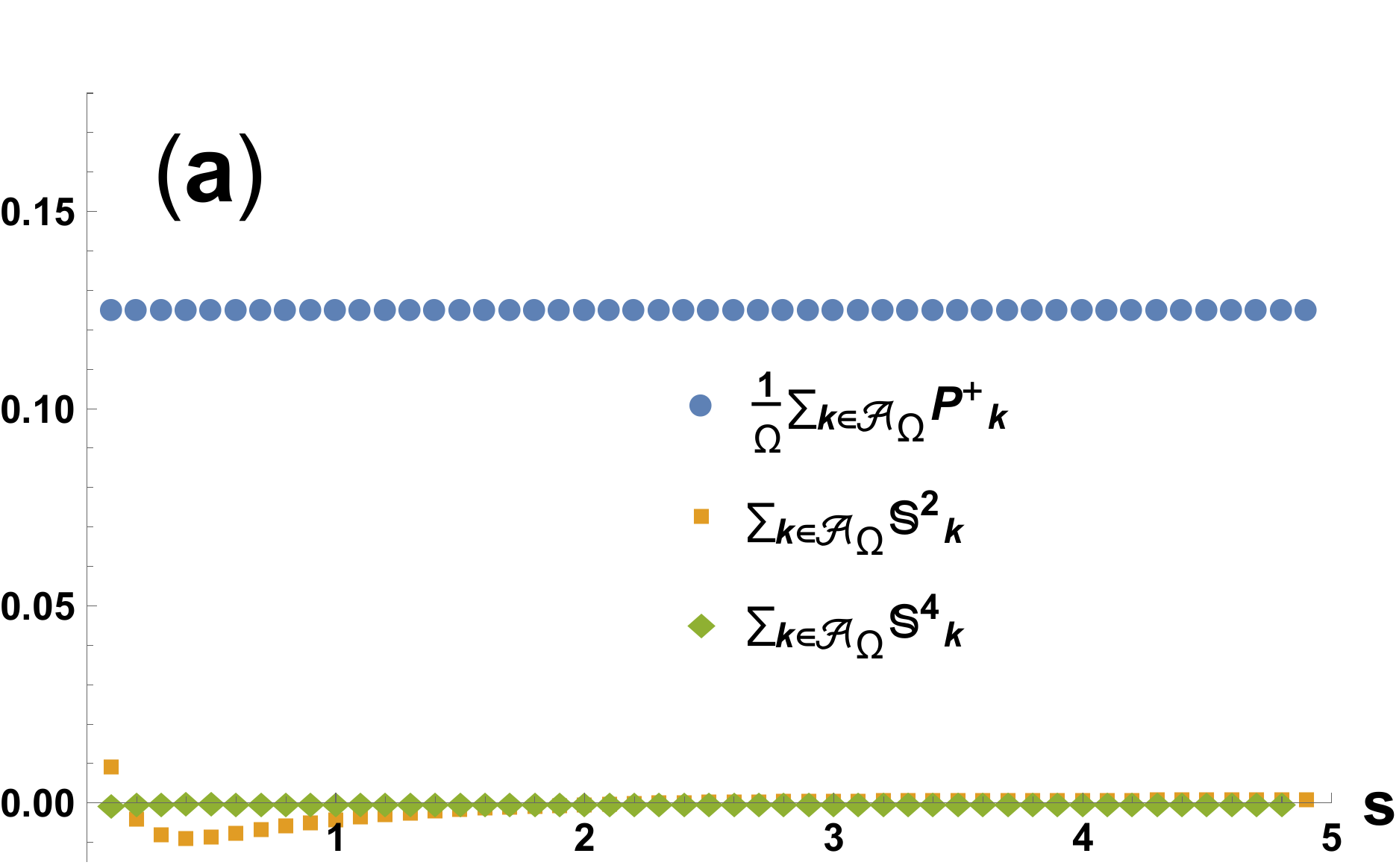}
 \includegraphics[width=0.48\textwidth]{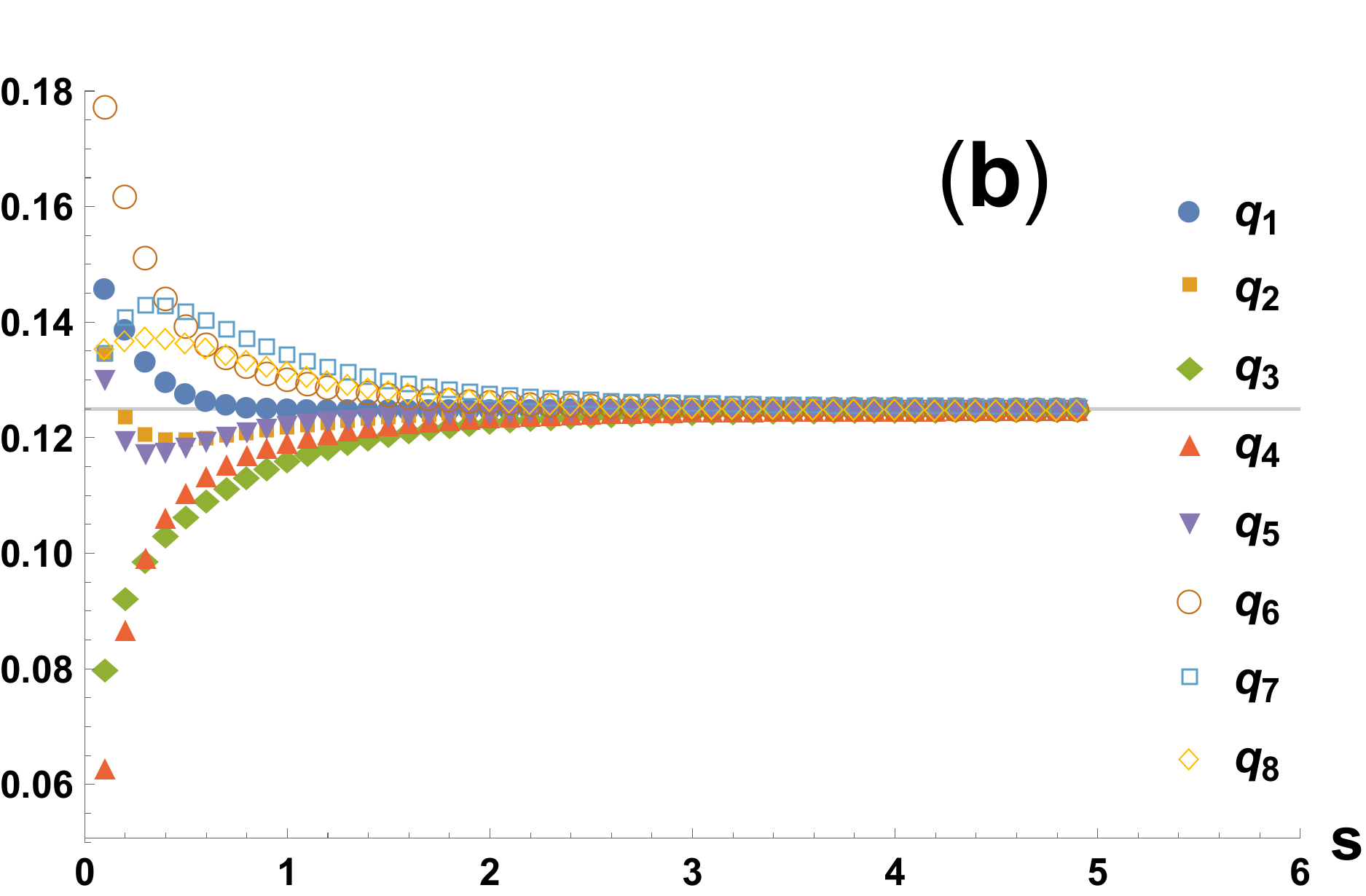}
 \includegraphics[width=0.48\textwidth]{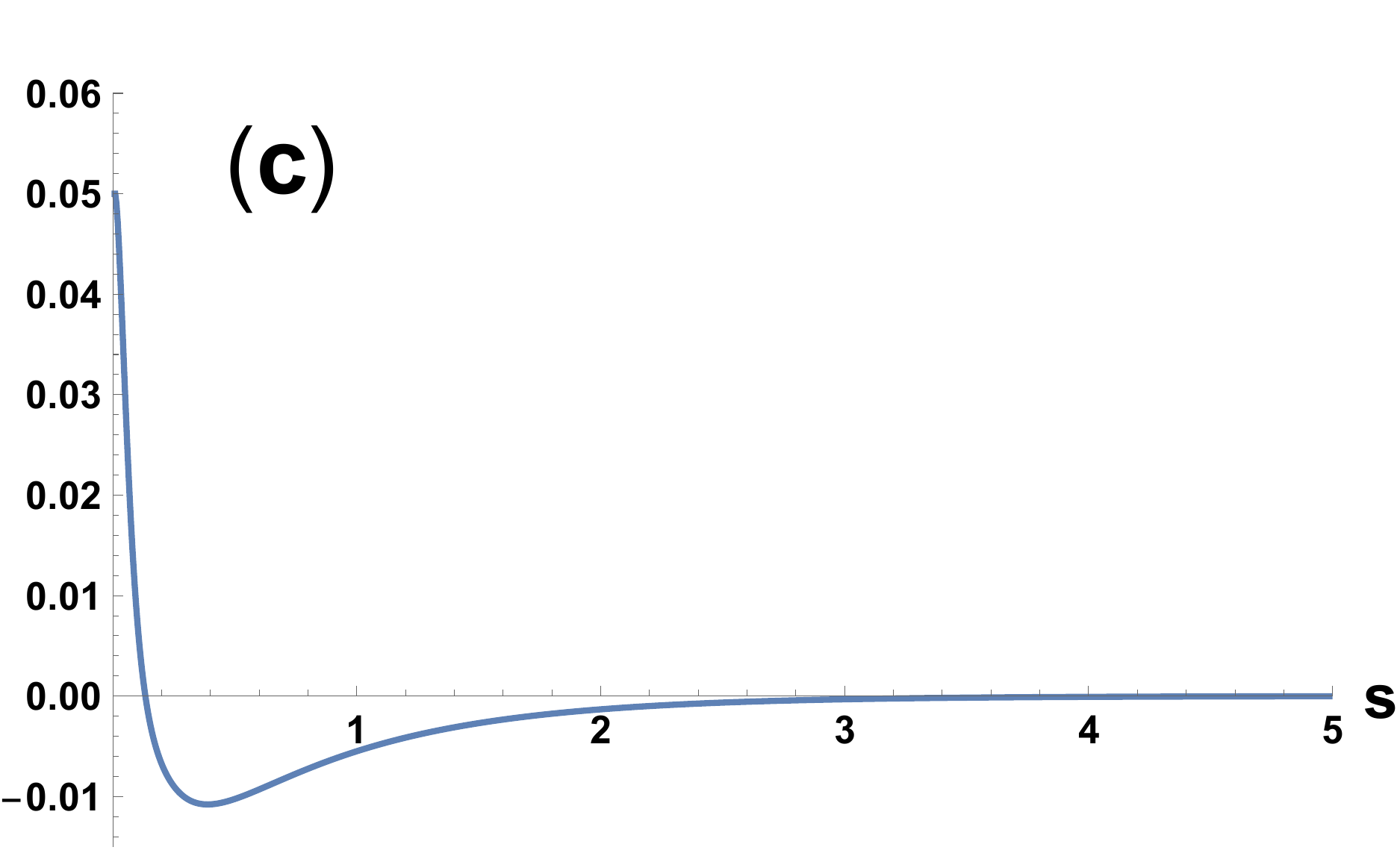}
\caption{\label{fig:Conservation_Axi}
 The plots are evaluated for $\eta = 10$.  (a) The expectation values for $\sum_{k\in \mathcal{A}_\Omega} P^+ _k$ and $\sum_{k\in \mathcal{A}_\Omega} \mathbb{S}^2 _k, \sum_{k\in \mathcal{A}_\Omega} \mathbb{S}^4 _k$ are evaluated. The conservation of $Q=1$ is validated (blue dots). (b) All the $q_k(s)$ values converge at the effective dynamics (also the steady state) to the  uniform value $q_k  \rightarrow 1/\Omega = 1/8$ represented by the gray horizontal line.  (c) The  $S_\xi  $ convergence to zero at the effective dynamics. }
\end{center}
\end{figure}

\begin{figure}
\begin{center}
 \includegraphics[width=0.48\textwidth]{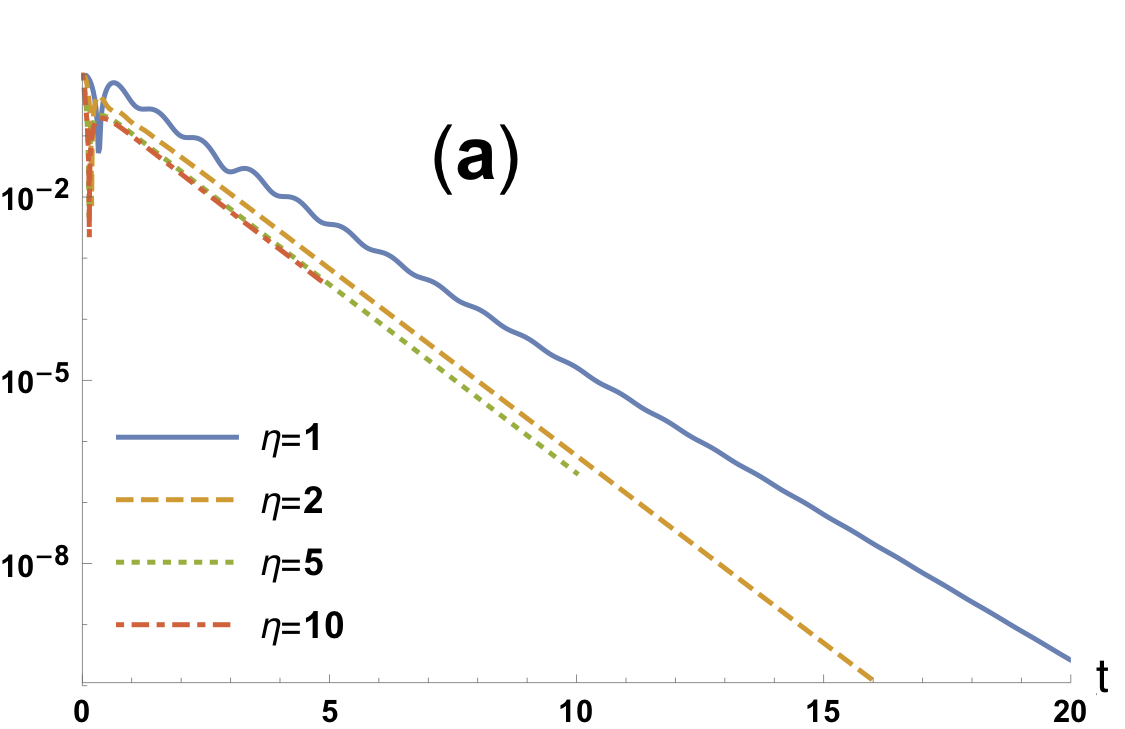}
 \includegraphics[width=0.48\textwidth]{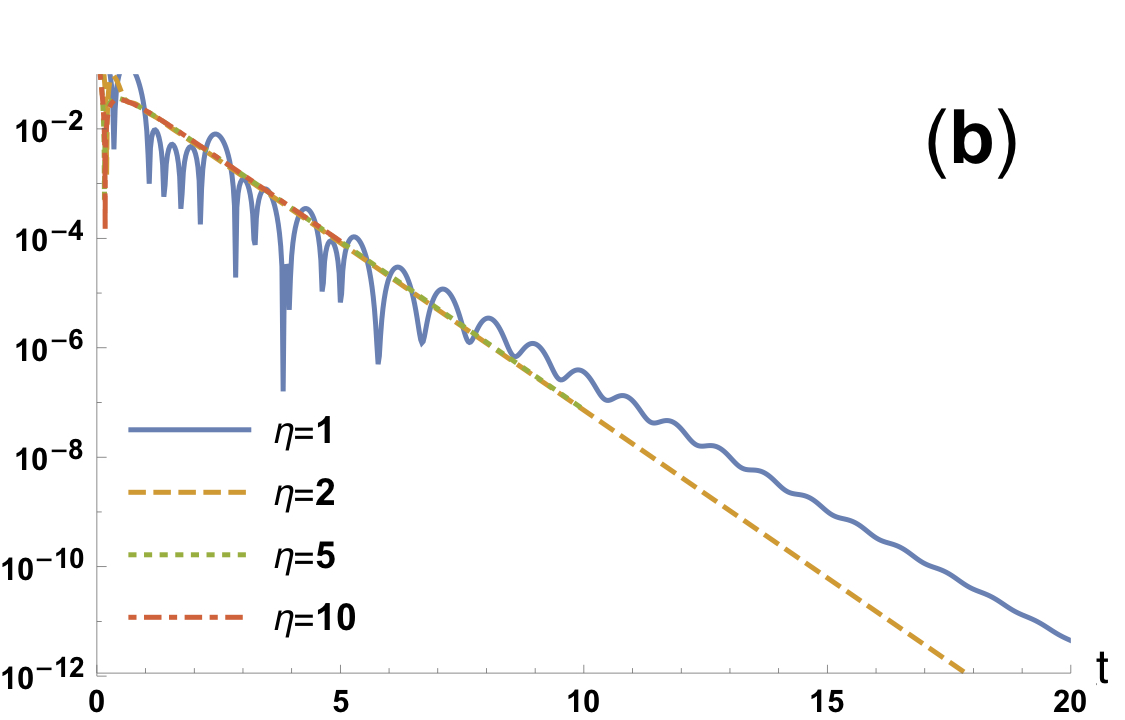}
 \includegraphics[width=0.48\textwidth]{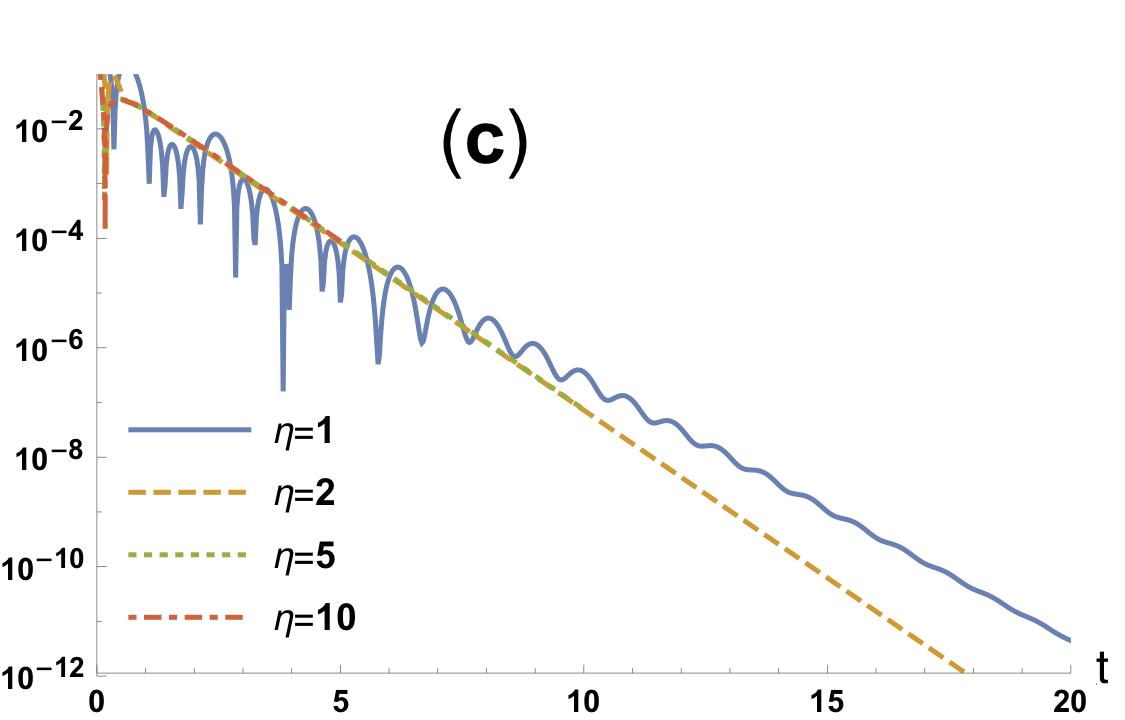}
 \includegraphics[width=0.48\textwidth]{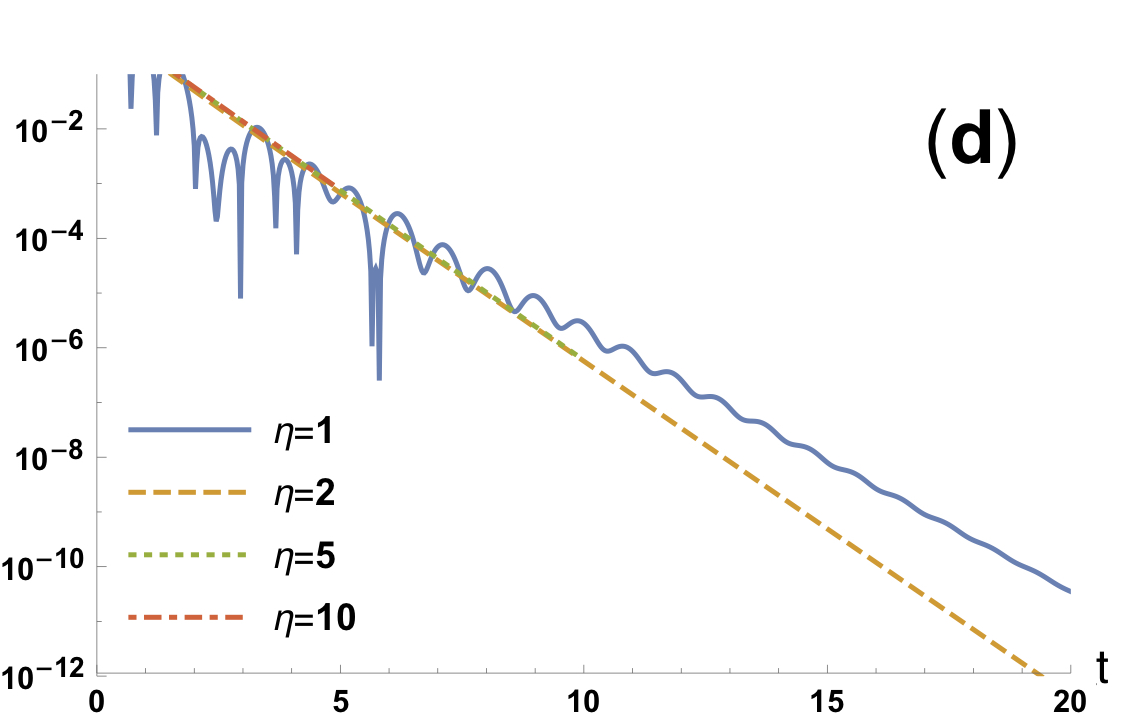}
\caption{\label{fig:Error_est_Axi}
The relative error estimation in logarithmic scale for (a) $S_\xi$ and (b-d) $q_{k=2,3,4}$, for   $\eta=\lbrace{1,2,5,10}\rbrace$ at the rescaled time $s=t/\eta$. The perturbation theory clearly converges at large $\eta$. Notice that the $y$ axis is given in logarithmic scale and that the $x$ axis represents the $t$ time rather than the rescaled $s=t/\eta$ time.  }
\end{center}
\end{figure}



\section{ Numerical support for the   $\mathcal{A_{\rm odd}}$ dephasing       }

The purpose of this section is to provide numerical support for the effective dynamic predictions of the $\mathcal{A}_{\rm odd}$ dephasing model at $Q=1$. The numerical protocol  directly evolves the  density matrix evolution according to equation (1) for the $\mathcal{A}_{\rm odd} $ selective dephasing model. We numerically evolve the density matrix for  $\Omega=8$ sites  with $\beta = 0.05, \epsilon=\nu_f=1,\eta=10$  and starting at the initial density matrix
\begin{equation}
\label{eq:initial conditions}
\rho_0 = \frac{1}{\Omega}\sum_{k\in \mathcal{A}_\Omega} \mathbb{P}_k + \beta ( \mathbb{S}^2 _2 +2\mathbb{S}^2 _4 - \mathbb{S}^2 _6 + \mathbb{P}_2- \frac{2}{3} \mathbb{P}_3  -\frac{1}{3}\mathbb{P}_7  ).
\end{equation}

We note that different initial conditions were tested producing similar results. To avoid repetitivity, we present only the results of the initial conditions \eqref{eq:initial conditions}.

\begin{figure}
\begin{center}
 \includegraphics[width=0.45\textwidth]{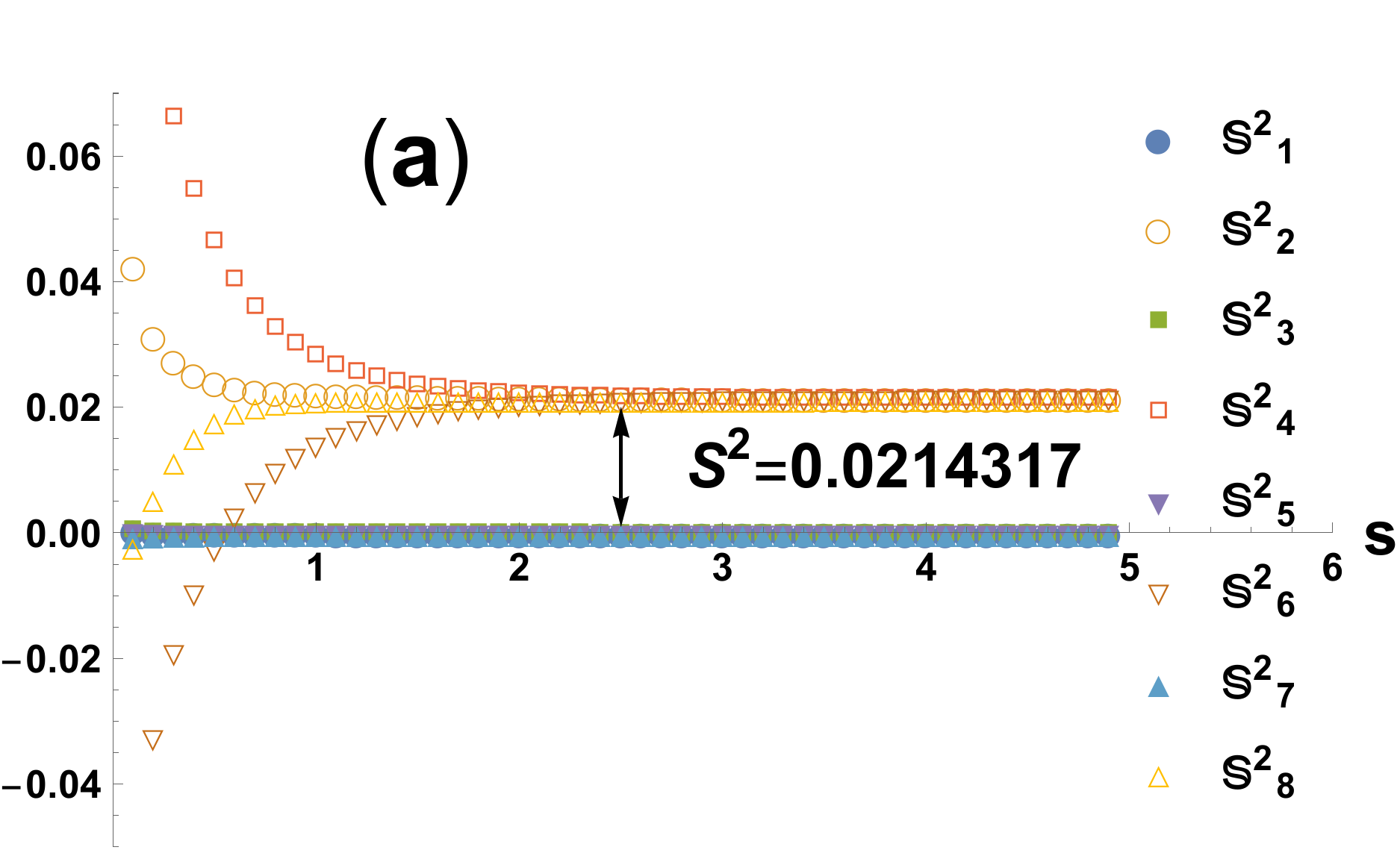}
 \includegraphics[width=0.45\textwidth]{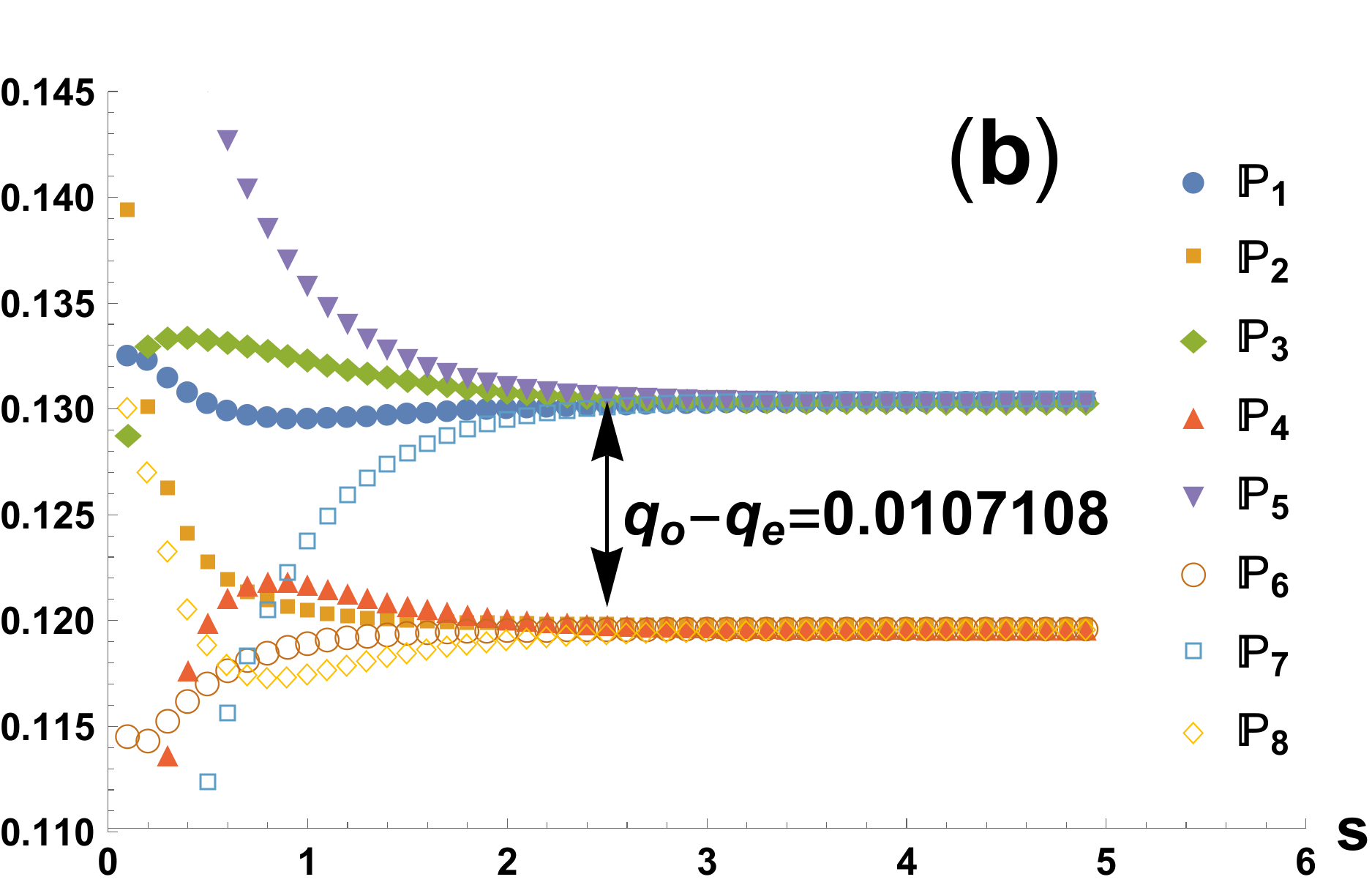}
  \includegraphics[width=0.45\textwidth]{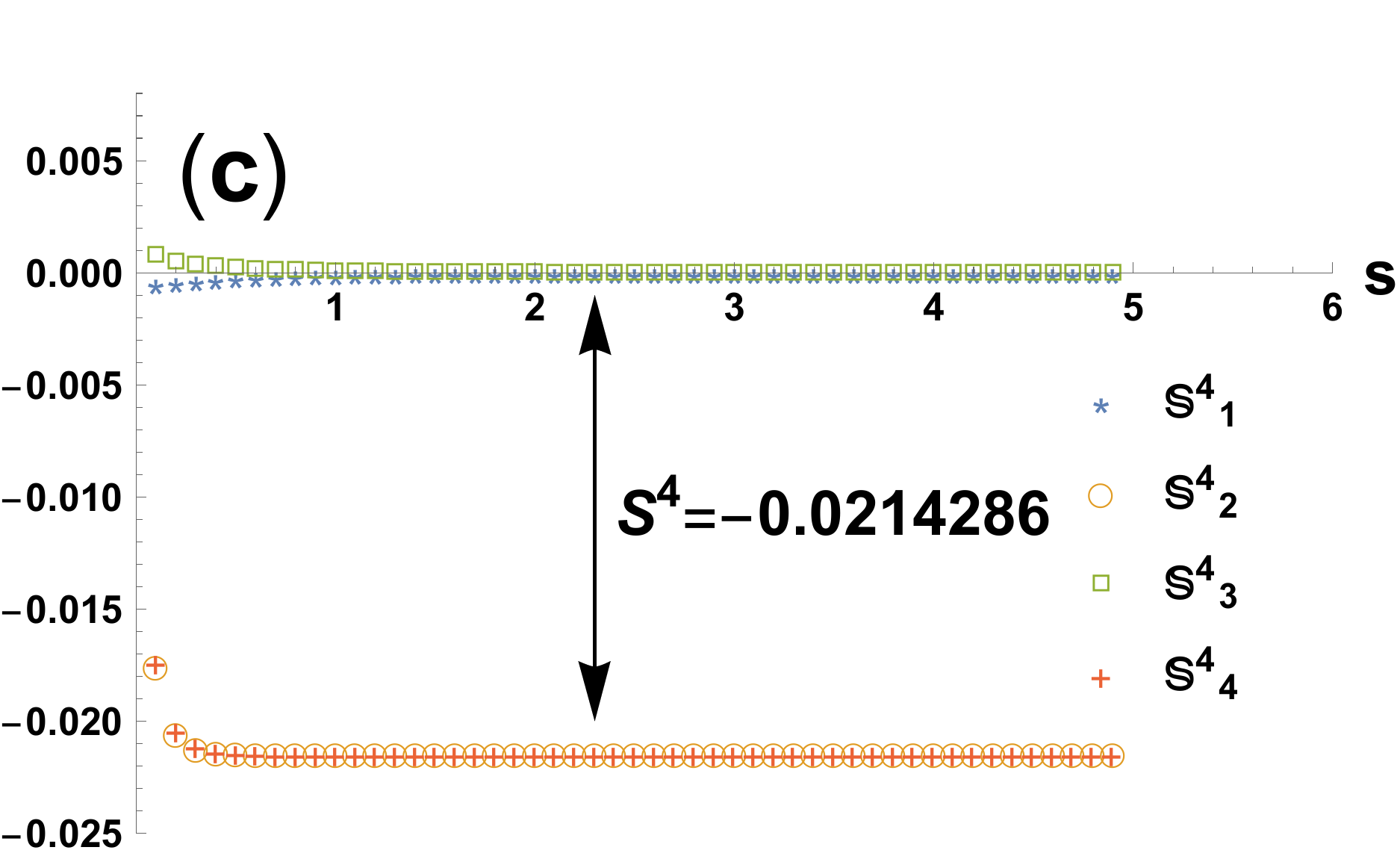}
\caption{\label{fig:EvoS2kPk_Aodd}
 Evaluation at $\eta=10$ for the $\mathcal{A}_{\rm odd}$ selective dephasing model in the rescaled time $s=t/\eta$. (a) Evolution of the expectation values of  $\mathbb{S}^2 _k $ for $k=1,...,8$. Only the even $k$ expectation values do not vanish, as $\mathbb{S}^2 _{2k+1} $ are not pointer states. (b) The evolution equation of the expectation of the operators $\mathbb{P}_k$. The system relaxes to its steady state around $s \sim 2$ which separates even from odd sites. (c) Evolution of the expectation values of  $\mathbb{S}^4 _k $ for $k=1,...,4$ as for $\Omega=8$, $\mathbb{S}^4 _{k+4} =\mathbb{S}^4 _k $. Only the even $k$ expectation values do not vanish, as $\mathbb{S}^4 _{2k+1} $ are not pointer states.
 At the steady state the prediction $q_{\rm o}-q_{\rm e} = \frac{1}{2}S^2 = - \frac{1}{2} S^4$ is verified to a good agreement. 
}
\end{center}
\end{figure}

In Fig.~\ref{fig:EvoS2kPk_Aodd}  we plot the evolution of the expectation values of $\mathbb{P} _k,\mathbb{S}^2 _k,\mathbb{S}^4 _k$. We qualitatively validate the steady state predictions $q_{\rm o}- q_{\rm e} = \frac{1}{2} S^2 = - \frac{1}{2} S^2 $ at the large time limit.  

\begin{figure}
\begin{center}
 \includegraphics[width=0.45\textwidth]{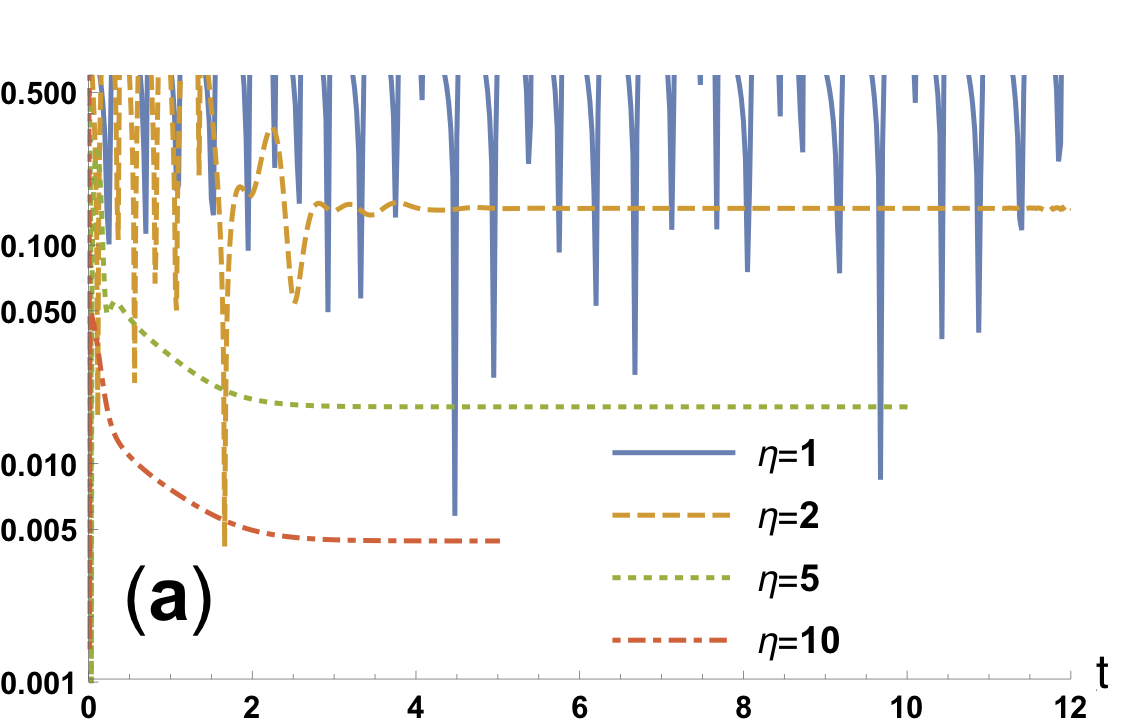}
 \includegraphics[width=0.45\textwidth]{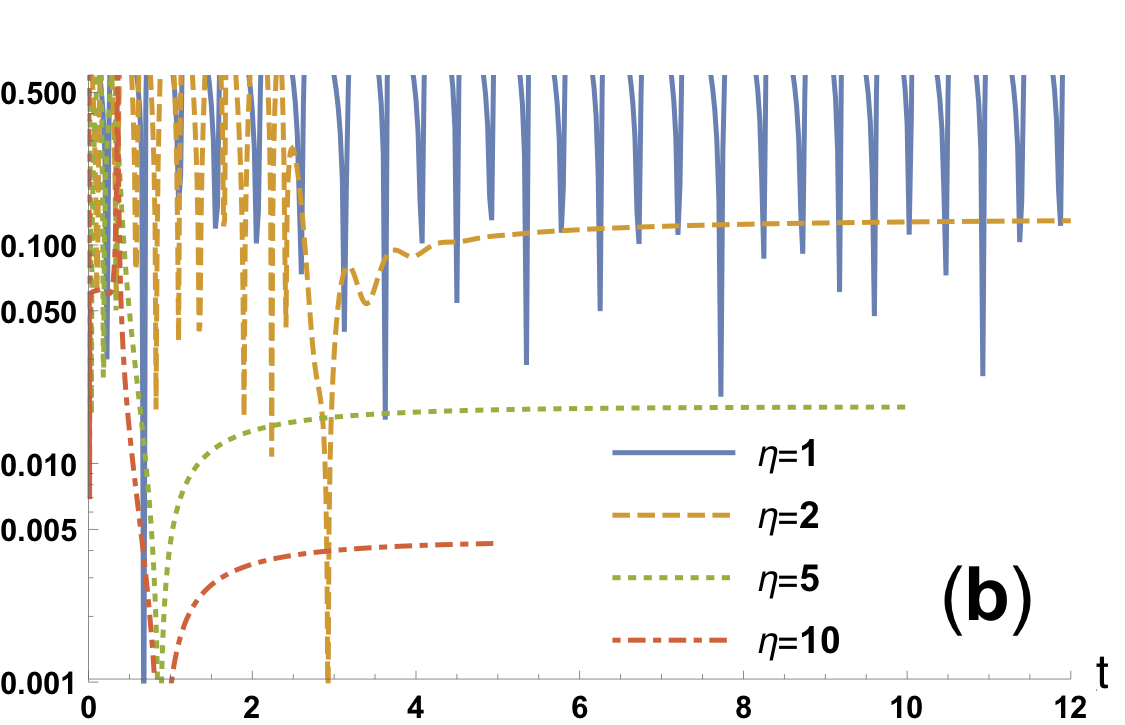}
 \includegraphics[width=0.45\textwidth]{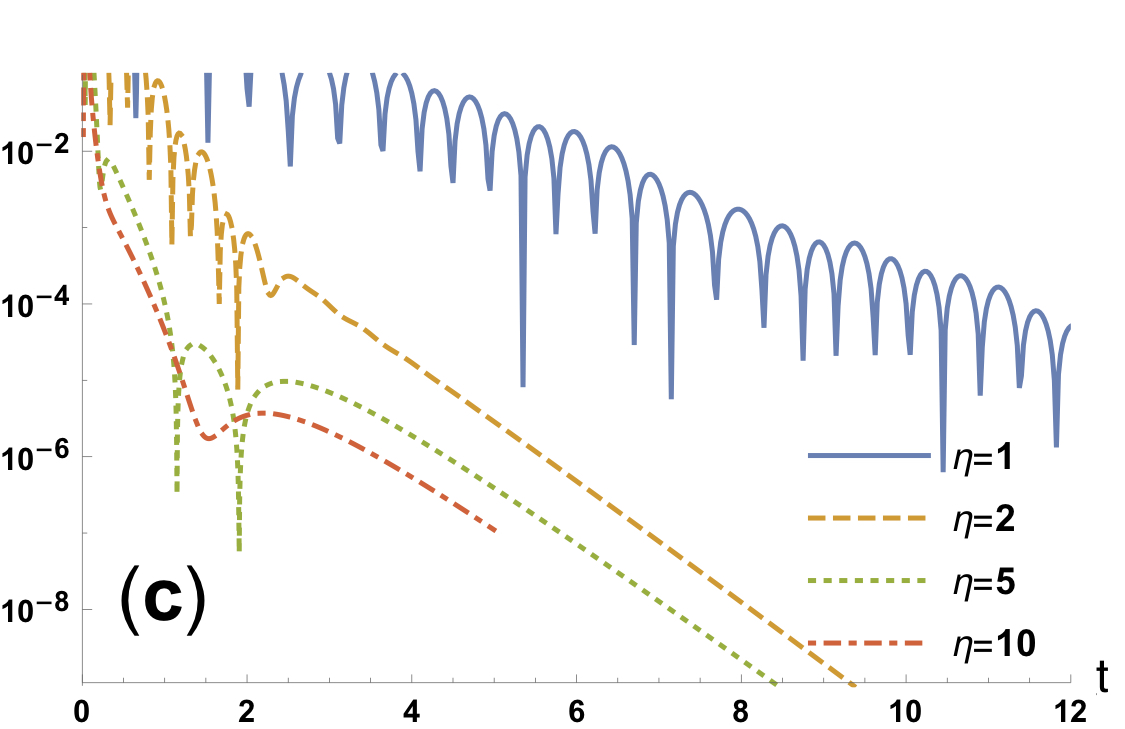}
 \includegraphics[width=0.45\textwidth]{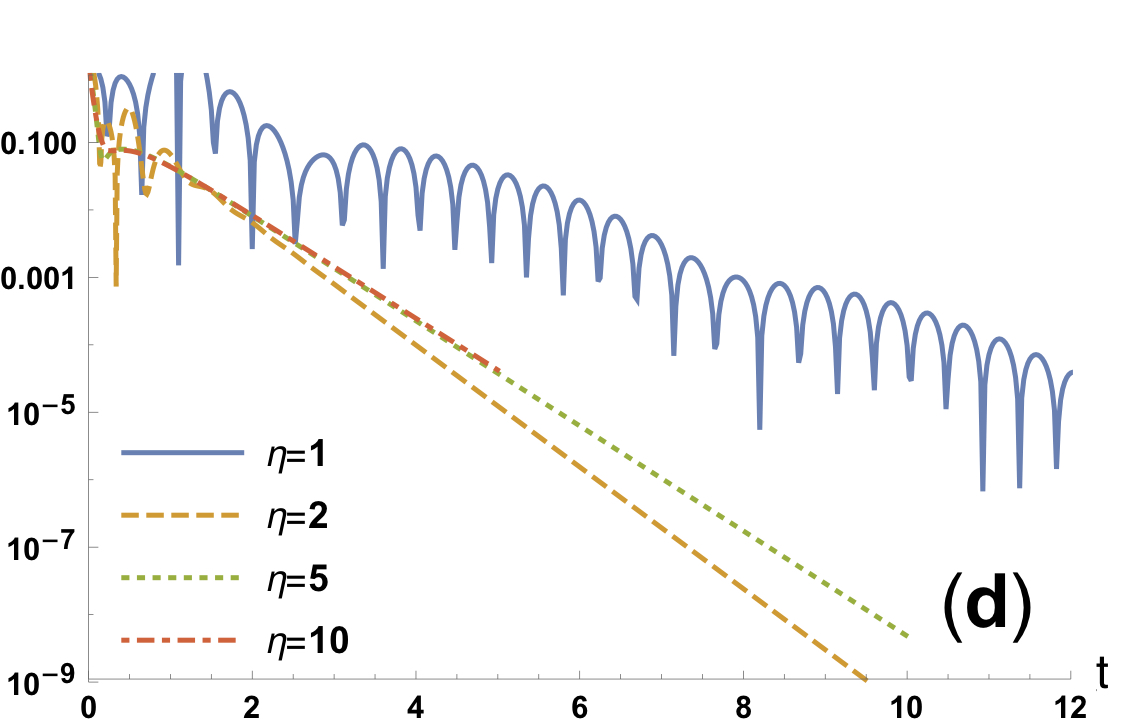}
 \includegraphics[width=0.45\textwidth]{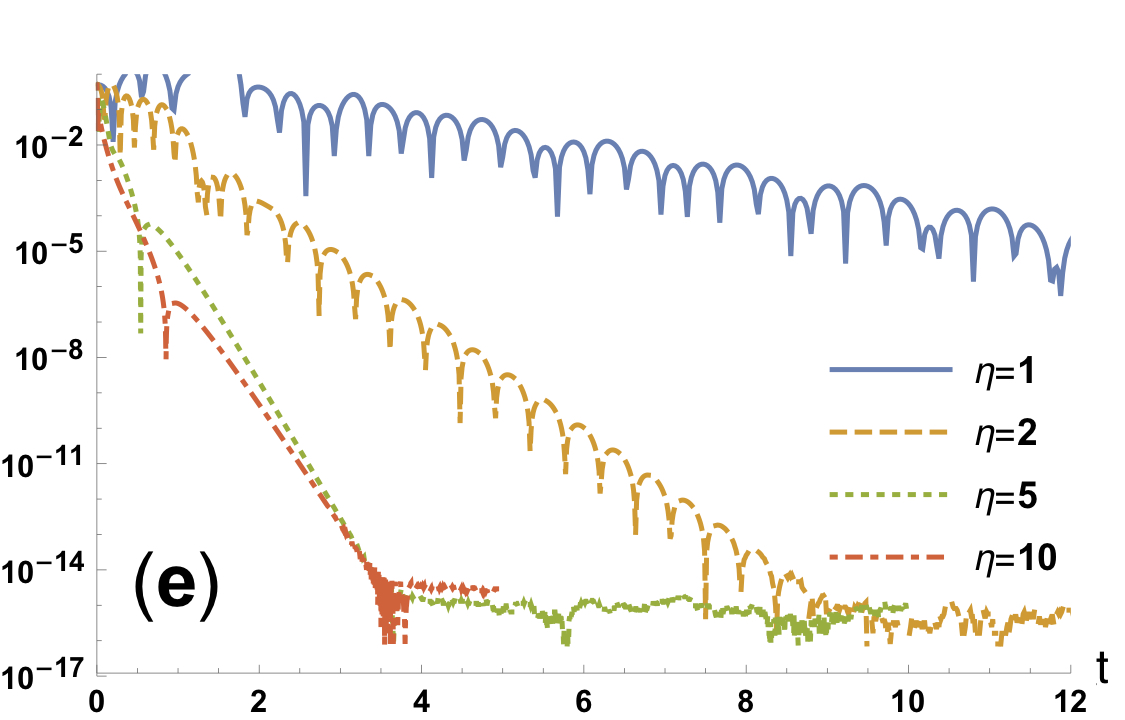}
 \includegraphics[width=0.45\textwidth]{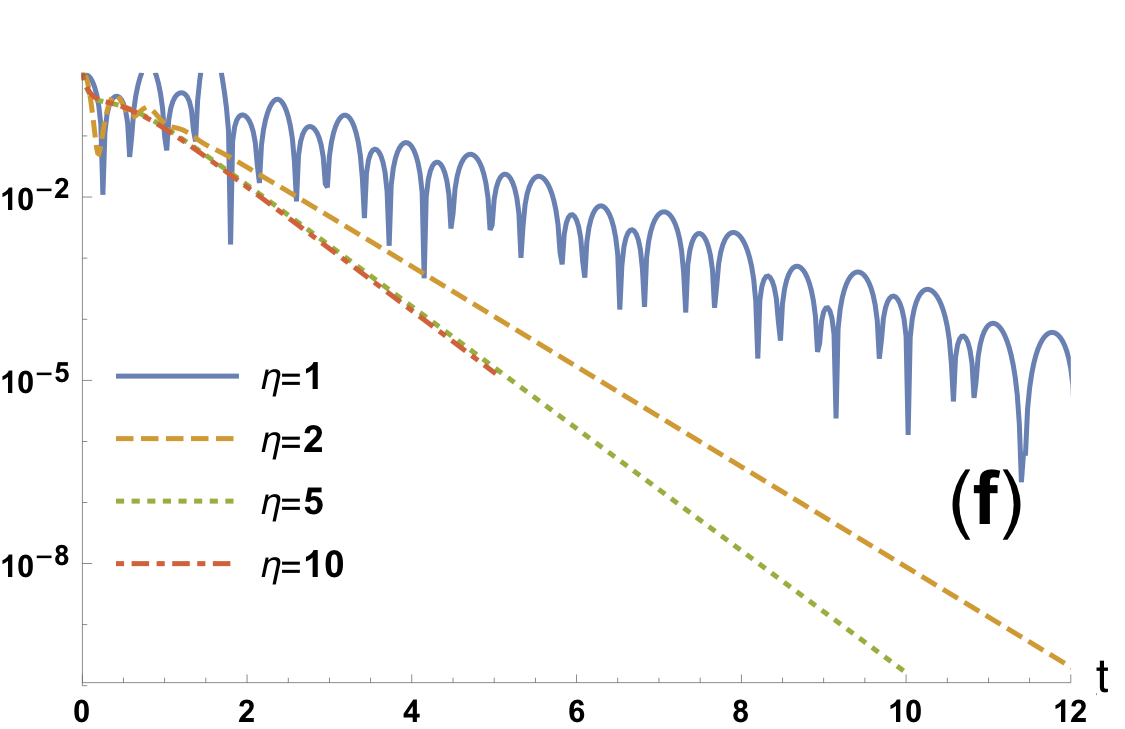}
\caption{\label{fig:Errors_Aodd}
 Evaluation of the error functions $E(s)$ for the equation (8), associated with  $\partial_s \mathbb{P}_{2,3}$ (a,b), $\partial_s\mathbb{S}^2 _{2,3}$ (c,d) and $\partial_s \mathbb{S}^4 _{2,3}$ (e,f)   in the $\mathcal{A}_{\rm odd}$ selective dephasing model with $\eta =1,2,5,10$. The  error vanishes as we increase $\eta$.  For $\eta=1$, there is poor convergence for the observed time scale. Already at $\eta\geq 2 $, the convergence is reasonable beyond an initial convergence time $t \sim 1$  as expected from the perturbation theory. Notice that the $y$ axis is given in logarithmic scale and that the $x$ axis represents the $t$ time rather than the rescaled $s=t/\eta$ time.
}
\end{center}
\end{figure}

To make a quantitative evaluation of the steady state predictions and to demonstrate the convergence with $\eta$ to the effective dynamics, we define again an error function $E(s)$. For an expectation $a(s) = b(s)$, the error function is defined to be $E(s) = \frac{|a(s)-b(s)|}{|a(s)+b(s)|}$. We expect $E(s)$ to vanish with increasing $\eta$ for $s\gtrsim
2$, where the effective dynamics kicks in.  
In Fig.~\ref{fig:Errors_Aodd}, the $E(s)$ functions for the evolution equations in (8) are presented. Excellent convergence already at $\eta=2$ is demonstrated.  One can also define an error function for the two steady state predictions $q_{\rm o}-q_{\rm e} = \frac{1}{2}S^2 ,q_{\rm o}-q_{\rm e} = - \frac{1}{2}S^4 $. We find that at the long time limit ($s\approx 5$ for $\eta=10$), the two error functions are of the order of $10^{-4}$. Namely, the convergence to the steady state predictions is well established for $\eta=10$.


 \section{ The Peres-Horodecki criterion
 }

In this section, we show that the density matrix corresponding to the large $\eta$ limit of the $\mathcal{A}_{\rm odd}$ supports bipartite entanglement. A similar analysis can show for the $\mathbb{A}_{\lbrace  \xi \rbrace}$ selective dephasing model, a non-vanishing $S_{\xi}$ value implies a bipartite entanglement at sites $\xi,\xi+2$.

We consider a density matrix of the form \begin{equation}
\label{eq:general density matrix}
    \rho  = \sum_k \alpha_k \mathbb{P}_k  + \sum_{k,m} \beta^m _k \mathbb{S}^{2m} _{2k}.
\end{equation}
Clearly, \eqref{eq:general density matrix} is not the most general form for a density matrix for the selective dephasing model. However, the evolution equations show that starting from this form and at the large $\eta$ limit, this form is kept with $\alpha_k ,\beta^m _k $ varying in time. Naturally, there are limitations for the values   of $\alpha_k ,\beta^m _k $ to keep the density matrix positive. Keeping the unity trace implies $\sum_k \alpha_k = 1$ and Hermitianity implies $\alpha_k, \beta^m _k  $ are real. Assume that the density matrix is indeed positive. Next, we choose two sites $2k',2k''$ and trace out the rest of the system (assume $k''>k'$). The reduced density matrix is of the form 
\begin{equation}
    \rho_{2k',2k''} = 
    a_1 P^- \otimes  P^- 
  + a_2 P^- \otimes  P^+ 
  + a_3 P^+ \otimes  P^-
  + b (\sigma^+ \otimes  \sigma^- + \sigma^- \otimes  \sigma^+), 
\end{equation}
where $a_1 = \sum_{k\neq 2k',2k''} \alpha_k ,a_2 = \alpha_{2k''},a_3=\alpha_{2k'}$ are real numbers associated to the values of $\alpha_k$ and $b=\beta^{k''-k'} _{k'}$. The positivity of the reduced system implies $a_2,a_2,a_3\geq 0$ as well as $a_2 a_3 \geq b^2$. The Peres-Horodecki criterion shows that if the partial transpose of the density matrix $\rho^P _{2k',2k''}$ has a negative eigenvalue, the density matrix is non-separable. Partial transposing with respect to either $2k'$ or $2k''$ gives 
\begin{equation}
    \rho^P _{2k',2k''} = 
    a_1 P^- \otimes  P^- 
  + a_2 P^- \otimes  P^+ 
  + a_3 P^+ \otimes  P^-
  + b (\sigma^+ \otimes  \sigma^+ + \sigma^- \otimes  \sigma^-). 
\end{equation}
It is easy to verify that for any real and non-vanishing $b$ value, the  partial transpose of the density matrix has a negative eigenvalue.  This implies that our system is bipartite entangled between any two sites $2k',2k''$ as long as $\vert \beta^{k''-k'} _{k'}\vert >0$. Therefore,  the $\beta^{k''-k'} _{k'}$ values, associated with the expectation values of $\mathbb{S}^{k''-k'} _{k'}$ are witnesses of entanglement as claimed in the main text.



 \section{ The effective dynamics for $\mathcal{A}_{\rm odd}$ dephasing model at $Q\neq1$     }

In this section we analyze the behaviour of the the $\mathcal{A}_{\rm{odd}}$ selective dephasing model at $Q\neq1$. As in the $Q=1$ case, we write down the effective dynamics for the $Q=2$ case. This leads to a set of diffusive-like evolution equations of the relevant pointer states. However, the amount of relevant pointer states for $Q=2$ grows significantly, making it cumbersome to analytically find the steady state expressions. To demonstrate that the coherences do not vanish also for $Q\neq1$, we employ a numerical evaluation starting at random initial conditions. We show that with the random initial condition, not only the coherent values associated with the $\mathbb{S}^{2k} _{2m}, \mathbb{S}^{4k} _{2m}$ survive in the steady state.

Consider the $Q=2$ case. The following operators allow to describe the evolution of the effective dynamics.
\begin{eqnarray}
\mathbb{M}_{x,y} &=& \mathbb{P}^+ _x  \mathbb{P}^+ _y  \prod_{j\neq x,y} \mathbb{P}^- _j
\\ \nonumber 
\mathbb{Z}^{2l} _{x,y} &=& (\sigma^+ _x \sigma^- _{x+l} + \sigma^- _x \sigma^+ _{x+l}) \mathbb{P}^+ _y \prod_{j\neq x,x+2l,y} \mathbb{P}^- _j
\\ \nonumber 
\mathbb{U}^{2l_1,2l_2} _{x,y} &=& (\sigma^+ _x \sigma^- _{x+2l_1} + \sigma^- _x \sigma^+ _{x+2l_1}) 
(\sigma^+ _y \sigma^- _{y+2l_2} + \sigma^- _y \sigma^+ _{y+2l_2})
\prod_{j\neq x,x+2l_1,y,y+2l_2} \mathbb{P}^- _j
\end{eqnarray}
    
    For $\mathbb{M}_{x,y}$, We consider only $x<y$ as $\mathbb{M}_{x,y}=\mathbb{M}_{y,x}$ and   $\mathbb{M}_{x,x}$ is undefined. The effective evolution equation for  $\mathbb{M}_{x,y}$ is given by 
\begin{eqnarray}
\label{eq:M evo equation}
\partial_s \mathbb{M}_{x,y} &=&
2D \Delta^{\mathbb{M}\rightarrow \mathbb{M}}  \mathbb{M}_{x,y} - D \Delta^{\mathbb{M}\rightarrow \mathbb{Z}^2} \mathbb{Z}^2 _{x,y},
\end{eqnarray}
with the following definitions of the discrete Laplace operators (valid for $\Omega>4$). 
\begin{eqnarray}
\Delta^{\mathbb{M}\rightarrow \mathbb{M}} \mathbb{M}_{x,y} = \left\{
                \begin{array}{ll}
\mathbb{M}_{x-1,x+1}-2\mathbb{M}_{x,x+1}+\mathbb{M}_{x,x+2} &  y=x+1\\
(\Delta_1 + \Delta_2) \mathbb{M}_{x,y}& y>x+1
                \end{array}
              \right.
              \\ 
              \Delta^{\mathbb{M}\rightarrow \mathbb{Z}^2} \mathbb{Z}^2 _{x,y} = \left\{
                \begin{array}{ll}
 \mathbb{Z}^2 _{x-2,x+1}+\mathbb{Z}^2 _{x-1,x}+\mathbb{Z}^2 _{x,x+1}+\mathbb{Z}^2 _{x+1,x}&  y=x+1\\
 \mathbb{Z}^2 _{x-2,x+2}-2\mathbb{Z}^2 _{x-1,x+2}-2\mathbb{Z}^2 _{x,x+1}+\mathbb{Z}^2 _{x+2,x}-2\mathbb{Z}^2 _{x+1,x}&  y=x+2\\
\Delta_1 \mathbb{Z}^2 _{x-1,y} + \Delta_2 \mathbb{Z}^2 _{y-1,x} & y>x+2
                \end{array}
              \right.
\end{eqnarray}
where we define  $\Delta_1 (\Delta_2)$ as the discrete Laplace operator acting on the first (second) index of the operator.  The evolution equation for $\mathbb{Z}^2 _{x,y}$ is given, for $y\neq x , x+2 $, by 

\begin{equation}
\label{eq: Z2 evo equation}
\partial_s \mathbb{Z}^2 _{x,y} =  
-2D \Delta^{\mathbb{Z}^2\rightarrow \mathbb{M}}  \mathbb{M}_{x,y}
+ 2D \Delta^{\mathbb{Z}^2\rightarrow \mathbb{Z}^2}    \mathbb{Z}^2 _{x,y}
- D \Delta^{\mathbb{Z}^2\rightarrow \mathbb{Z}^4} \mathbb{Z}^4 _{x,y}
- D \Delta^{\mathbb{Z}^2\rightarrow \mathbb{U}^{2,2}} \mathbb{U}^{2,2} _{x,y}, 
\end{equation}
with the following definitions for the discrete Laplace equations. 
\begin{eqnarray}
\Delta^{\mathbb{Z}^2\rightarrow \mathbb{M}} \mathbb{M}_{x,y} &=& \left\{
\begin{array}{ll}
\mathbb{M}_{x-1,x}-2\mathbb{M}_{x-1,x+1}+\mathbb{M}_{x-1,x+2} &  y=x-1\\
\mathbb{M}_{x,x+1}-2\mathbb{M}_{x,x+2}+\mathbb{M}_{x+1,x+2} &  y=x+1\\
\Delta_1 \mathbb{M}_{x+1,y} & y\neq x\pm 1
\end{array}
              \right.
\end{eqnarray}
\begin{eqnarray}
\Delta^{\mathbb{Z}^2\rightarrow \mathbb{Z}^2} \mathbb{Z}^2 _{x,y} &=& \left\{
\begin{array}{ll}
-\mathbb{Z}^2 _{x-1,x}+(1+2\delta_{x,\rm{odd}})\mathbb{Z}^2 _{x,x-2}-(3+2\delta_{x,\rm{odd}})\mathbb{Z}^2 _{x,x-1}+\mathbb{Z}^2 _{x+1,x-1}
&  y=x-1\\
-\mathbb{Z}^2 _{x-1,x+2}-2\mathbb{Z}^2 _{x,x+1}-\mathbb{Z}^2 _{x+1,x}&  y=x+1\\
\mathbb{Z}^2 _{x-1,x+3}-(3+2\delta_{x,\rm{odd}})\mathbb{Z}^2 _{x,x+3}+(1+2\delta_{x,\rm{odd}})\mathbb{Z}^2 _{x,x+4}-\mathbb{Z}^2 _{x+1,x+2}&  y=x+3\\
(\Delta_1 +\Delta_2 +2 \delta_{k,\rm{odd}}\Delta_2)\mathbb{Z}^2 _{x,y} & y\neq x\pm 1,x+3
\end{array}
              \right.
               \\
\Delta^{\mathbb{Z}^2\rightarrow \mathbb{Z}^4} \mathbb{Z}^4 _{x,y} &=& \left\{
\begin{array}{ll}
-2 \mathbb{Z}^4_{x-2,x-1} \left(1+\delta _{k,\rm{odd}}\right)+\mathbb{Z}^4_{x-2,x} \left(1+2 \delta _{x,\rm{odd}}\right)-2 \mathbb{Z}^4_{x-1,x-2}+\mathbb{Z}^4_{x,x-2}&  y=x-2\\
-2 \mathbb{Z}^4_{x-2,x} \left(1+\delta _{x,\rm{odd}}\right)+\mathbb{Z}^4_{x-2,x-1} \left(1+2 \delta _{x,\rm{odd}}\right)+2 \mathbb{Z}^4_{x-1,x}+\mathbb{Z}^4_{x,x-1}&  y=x-1\\
-\mathbb{Z}^4_{x,x+2} \left(2+2 \delta _{k,\rm{odd}}\right)+\mathbb{Z}^4_{x,x+3} \left(1+2 \delta _{k,\rm{odd}}\right)+\mathbb{Z}^4_{x-2,x+3}+2 \mathbb{Z}^4_{x-1,x+2}&  y=x+3\\
-2 \mathbb{Z}^4_{x,x+3} \left(1+\delta _{x,\rm{odd}}\right)+\mathbb{Z}^4_{x,x+2} \left(1+2 \delta _{k,\text{odd}}\right)+\mathbb{Z}^4_{x-2,x+4}-2 \mathbb{Z}^4_{x-1,x+4}&  y=x+4\\
\Delta_1  \mathbb{Z}^4 _{x-1,y}& \rm{else}
\end{array}
              \right.
                \\
\Delta^{\mathbb{Z}^2\rightarrow \mathbb{U}^{2,2}} \mathbb{U}^{2,2} _{x,y} &=& \left\{
\begin{array}{ll}
0& y-x=\pm1 , 3 \\
\mathbb{U}^{2,2} _{x,x-4}& y-x=-2\\
\mathbb{U}^{2,2} _{x,x+4}&  y-x=4\\
\Delta_2 \mathbb{U}^{2,2} _{x,y-1}& \rm{else}
\end{array}
              \right.
\end{eqnarray}
As in the $Q=1$ case, there are emerging conservation rules in the effective dynamics, namely 
\begin{eqnarray}
 \partial_s \sum_{x,y} \mathbb{M}_{x,y}  =0, & \quad\quad&
 \partial_s \left(-2\mathbb{Z}^2 _{x,x+1}+ \sum_{x,y} \mathbb{Z}^2 _{x,y} \right) =0.
\end{eqnarray}
Here we have assumed no counting over the undefined states $\mathbb{M}_{x,x}, \mathbb{Z}^2_{x,x},\mathbb{Z}^2_{x,x+2}$. While the conservation associated with  $\mathbb{M}_{x,y}$ is valid even for the $t$ dynamics,  the conservation associated with $\mathbb{Z}^2 _{x,y}$ is valid only for the effective dynamics, similarly to the $Q=1$ case with $\mathbb{P}_x$ and $ \mathbb{S}^2 _{x}$ respectively. 

Solving even the steady state equations of the effective dynamics for some given $\Omega$ sites requires writing down the evolution equations for more states, e.g. $\mathbb{U}^{2l_1,2l_2} _{x,y}, \mathbb{Z}^4 _{x,y}$. This will not be carried out here. However, similar to the $Q=1$ case, we see that uniform values at the steady state of the diagonal terms  $\mathbb{M}_{x,y}$ leads to vanishing coherences. We thus conjecture that for any $Q$, the scale of the surviving coherent terms, e.g. $\mathbb{S}^{2m} _{2k}, \mathbb{Z}^{2m} _{2k_1,k_2}$, is proportional to the imbalance of their relevant diagonal terms.    

To support this conjecture we numerically evolve a density matrix with random initial conditions. Namely, we use the Cholesky decomposition and randomise a matrix $\mathcal{J}$ with values in the complex square $\lbrace -1-i,1+i \rbrace $. Then, the (positive semidefinite, Hermitian and trace $1$) initial density matrix is set to be  $\rho_0 = \frac{\mathcal{J}\mathcal{J}^\dagger}{\Tr \mathcal{J}\mathcal{J}^\dagger}  $. The diagonal states are expected to be close to the uniform case, namely to have a value close to $2^{-\Omega}$. To highlight the offset from this average and compare it with the surviving coherences, we define the matrix $\varrho$, with the matrix elements $\varrho_{ij} = |\rho_{ij}(s) - 2^{-\Omega} \delta_{ij}|$. We aim to show that at the steady state, the diagonal states are of the same order as the surviving off-diagonals (coherences).

The advantage of this method is that it allows to confirm that the surviving coherences extend beyond the limited case $Q=1$, without meticulously following the effective dynamics evolution equations. There are two significant disadvantages. First, we note that since we start from a random initial condition the effect of the coherences would generally be weak. The effect cannot be tuned as was done through $\beta$ for $Q=1$ (see the inset of Fig.1). Secondly, it is hard to associate the relevant coherent pointer states to their respective diagonal pointer states. This implies that it is hard to learn which $Q$ is associated to a long lasting coherence term. 

\begin{figure}
\begin{center}
 \includegraphics[width=0.38\textwidth]{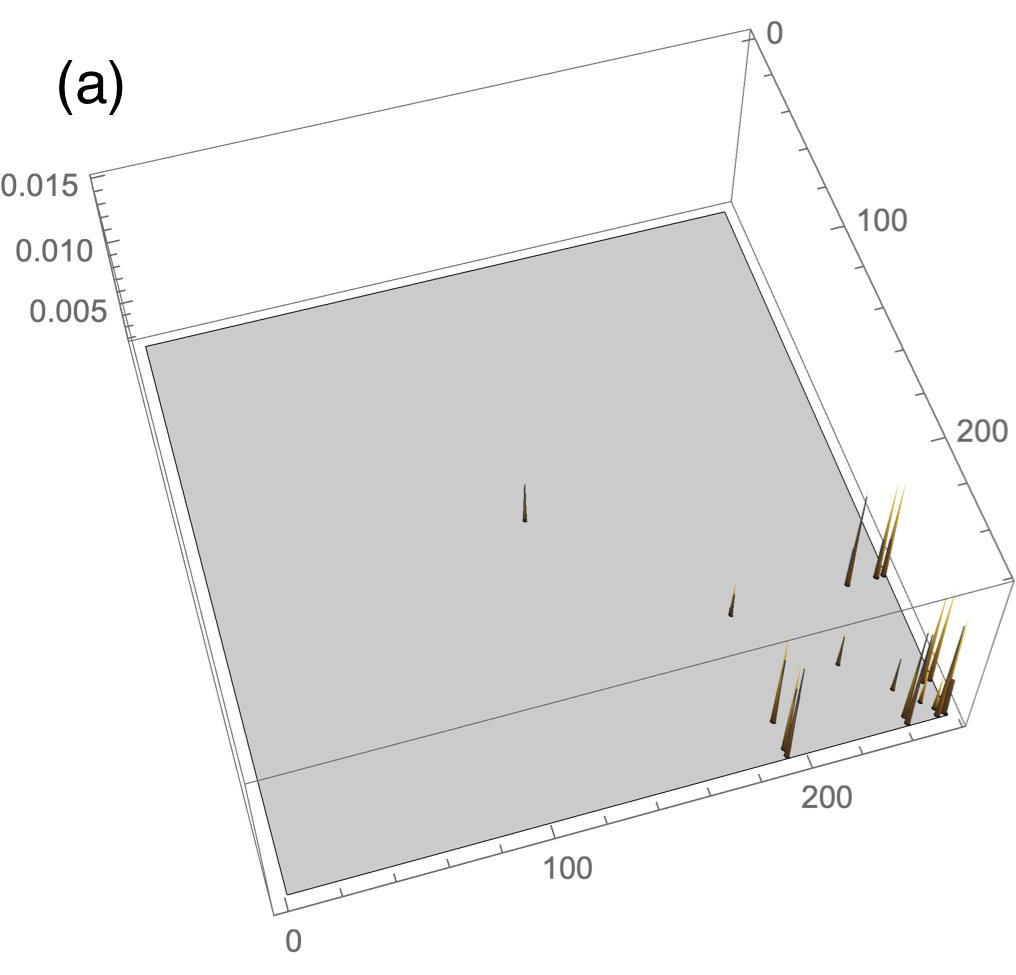}
 \includegraphics[width=0.46\textwidth]{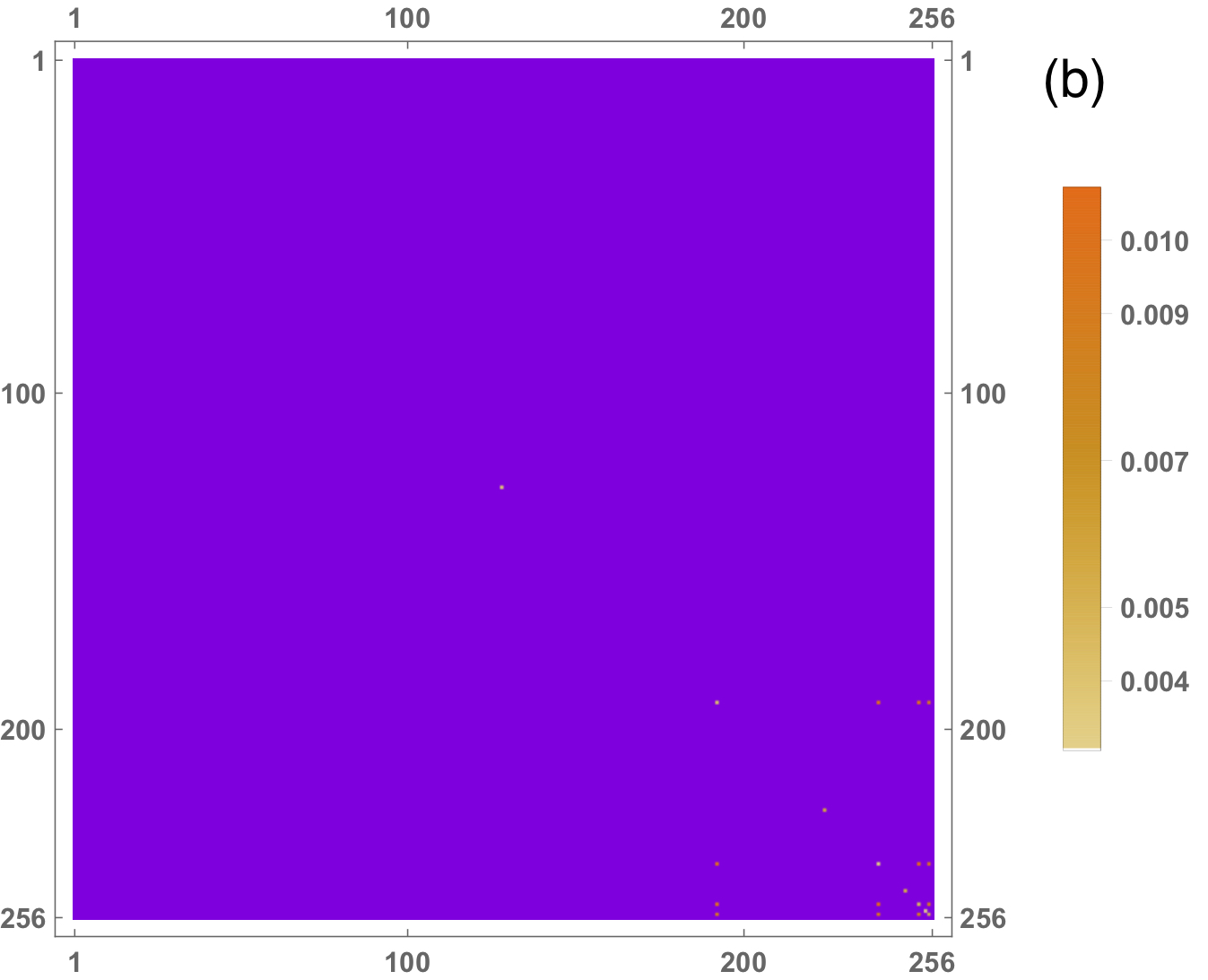}
 \includegraphics[width=0.38\textwidth]{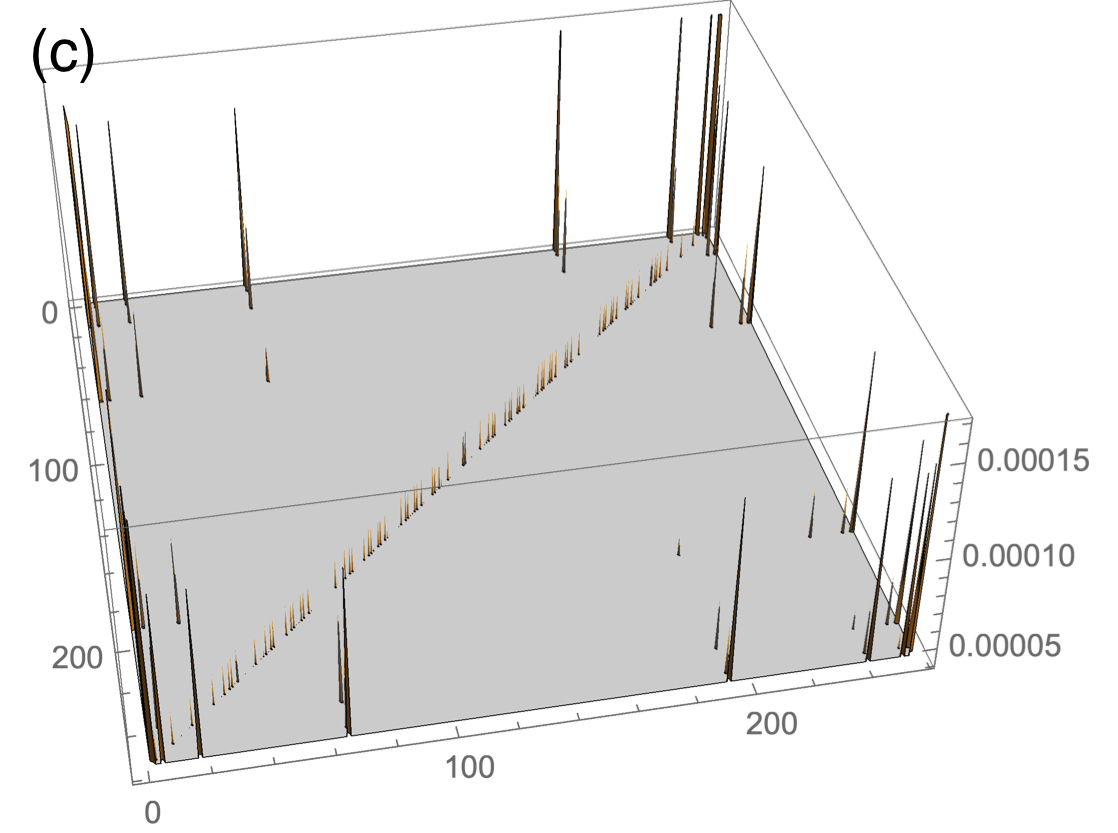}
 \includegraphics[width=0.46\textwidth]{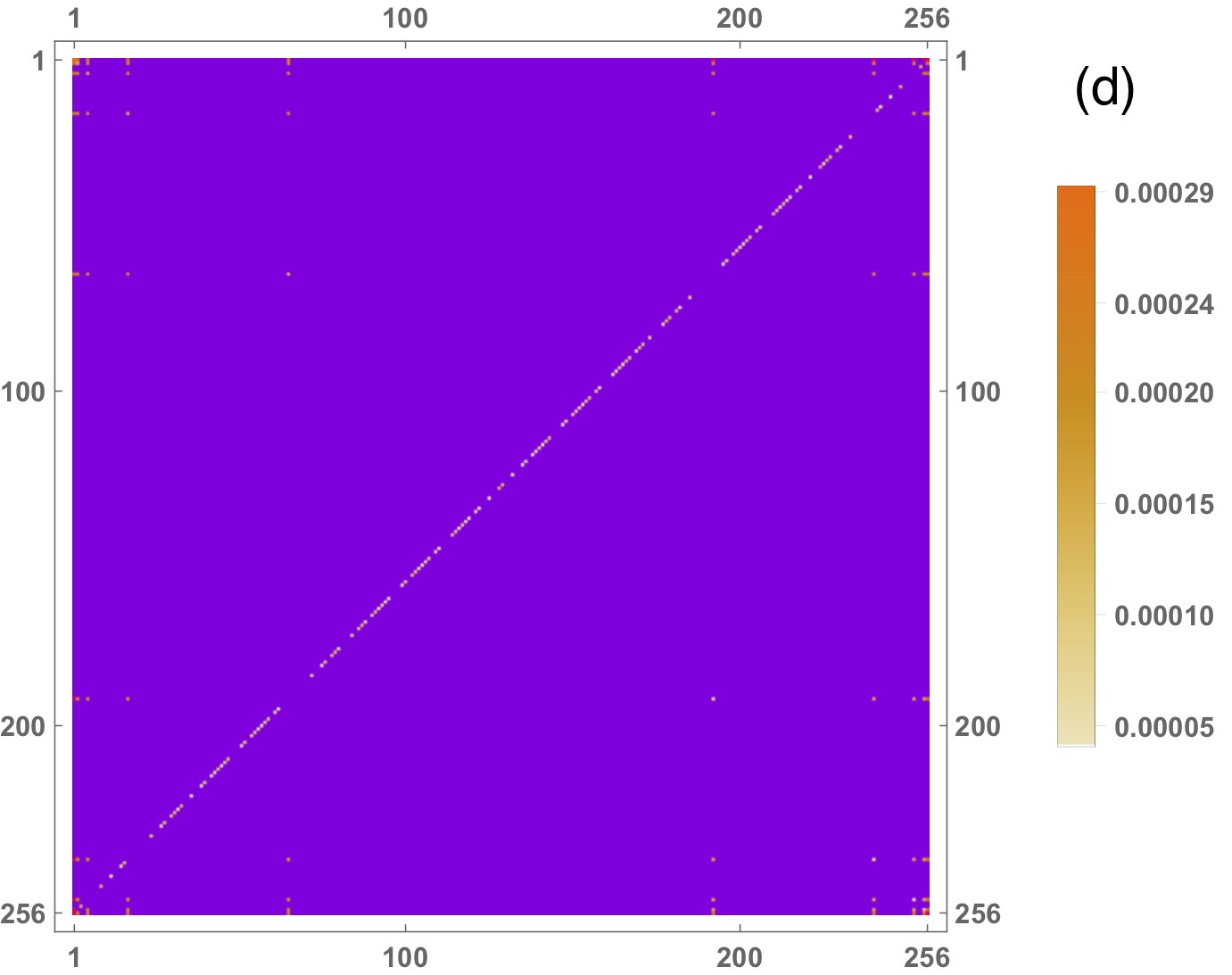}
 \includegraphics[width=0.38\textwidth]{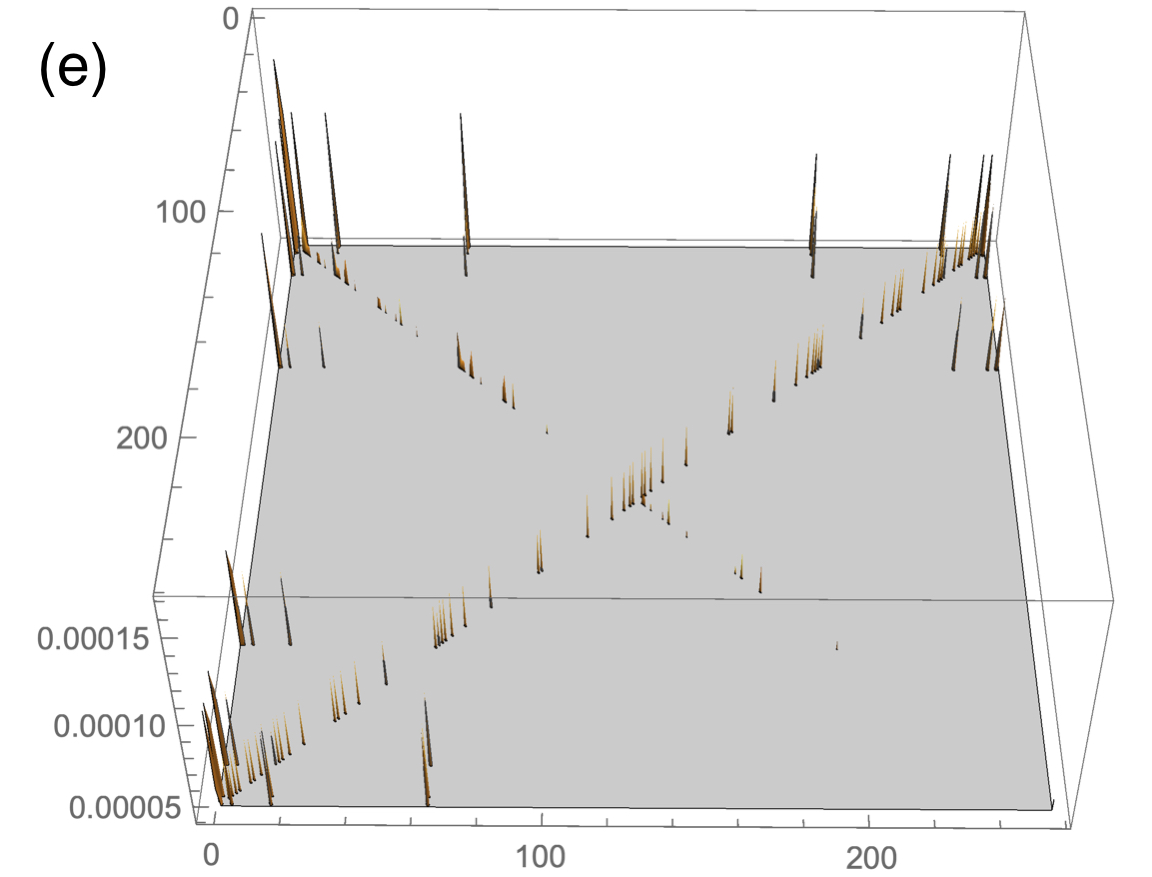}
 \includegraphics[width=0.46\textwidth]{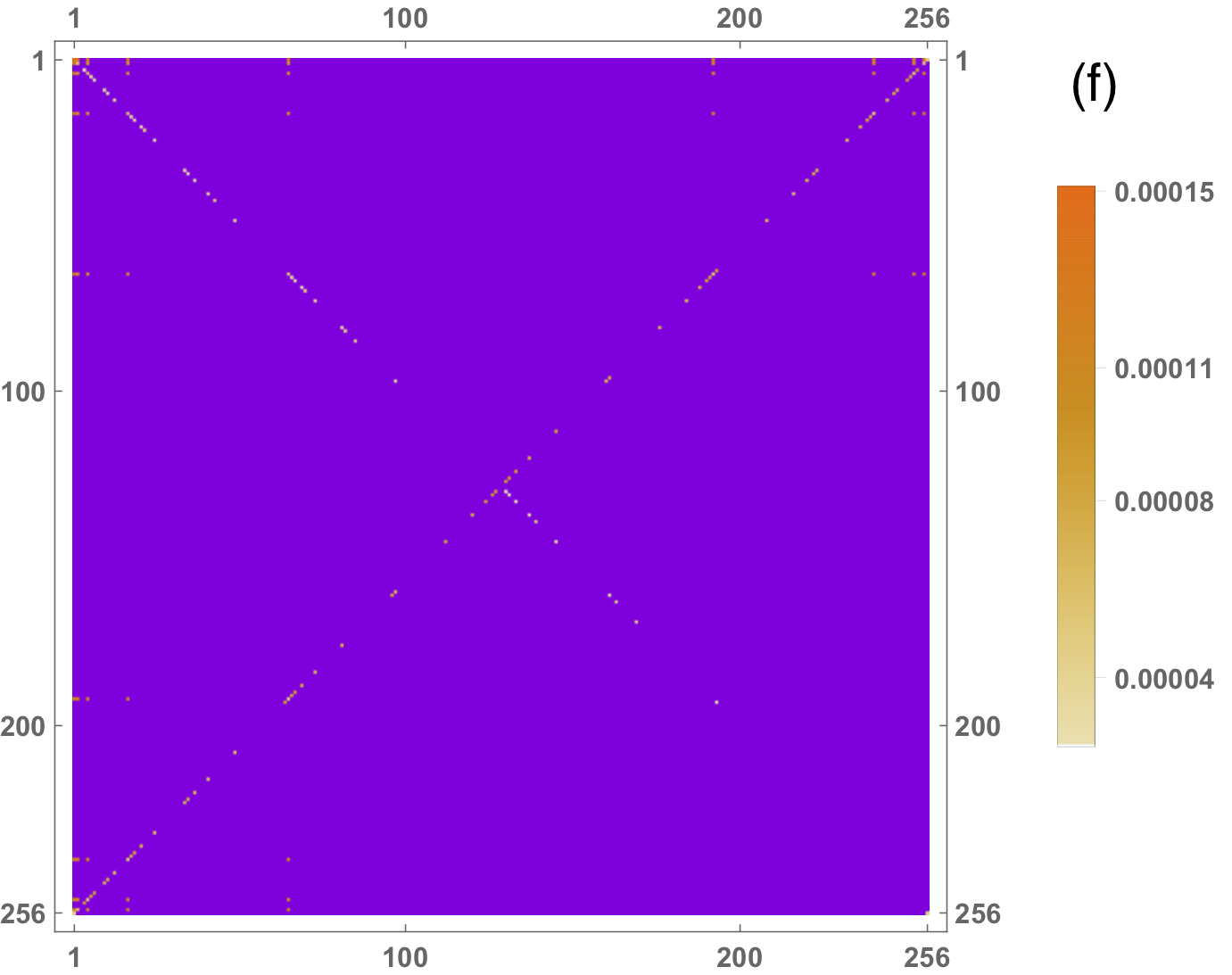}
\caption{\label{fig:rho_offset}
The offset density matrix $\varrho$ at $\eta=10$ for: (a,b) The $Q=1$ initial conditions \eqref{eq:initial conditions} with $\beta=0.05$. (c,d) The first realization of the random initial condition (e,f) The second realization of the random initial condition. In the first realization, the surviving coherent terms include those of $Q=1$, but in the second realization they are too weak to notice due to the random initial condition they obtained. 
}
\end{center}
\end{figure}

However, to demonstrate the effect of surviving coherences of the order of the diagonals' offset, the random initial conditions is adequate. In Fig.~\ref{fig:rho_offset} we show the matrix $\varrho$ for $\eta=10$, comparing between the initial conditions of $Q=1$ similarly to the main text ($\beta=0.05$) and two random initial conditions. It can clearly be seen that many more coherent terms appear for the random case. Moreover, as expected, the coherent terms are of the order of the diagonals' offset values.

\begin{figure}
\begin{center}
 \includegraphics[width=0.48\textwidth]{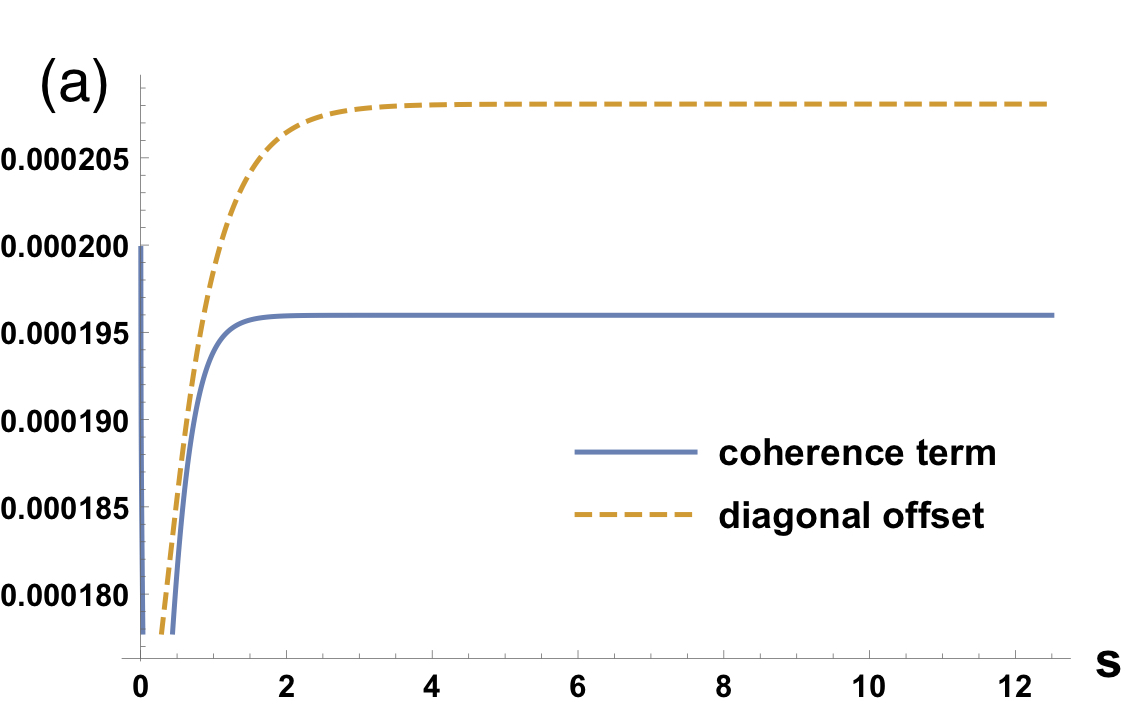}
 \includegraphics[width=0.48\textwidth]{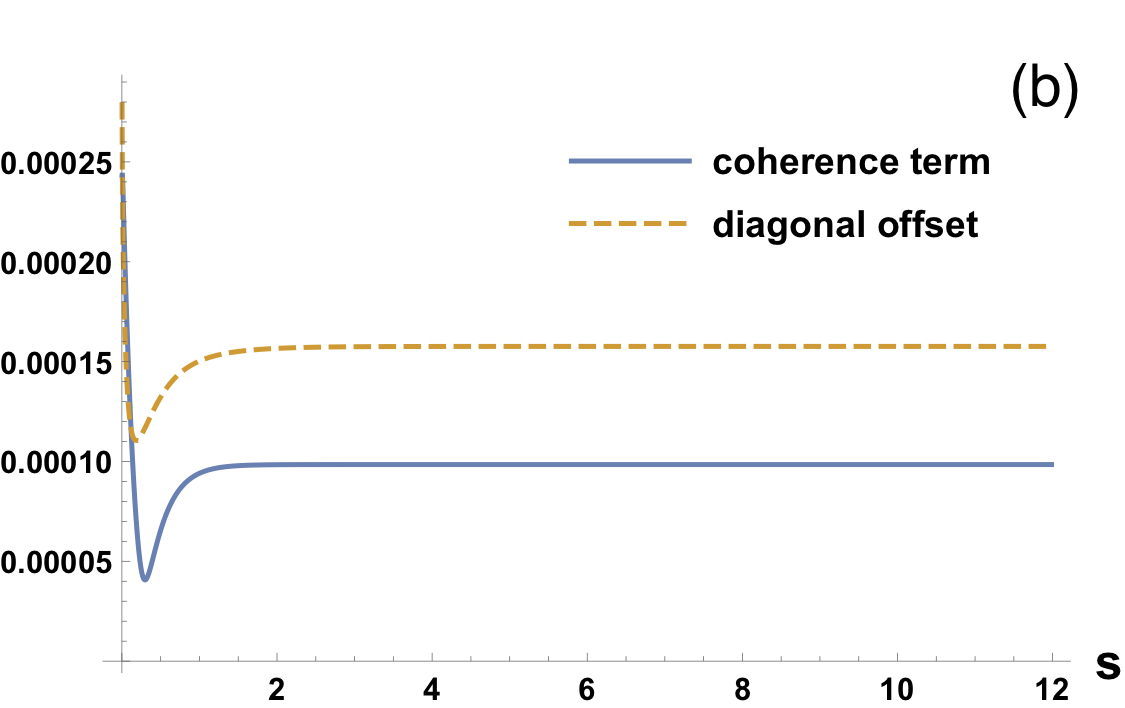}
\caption{\label{fig:coherences_steadystate}
(a,b) Evolution of the largest surviving coherent term $\rho(i,j)$ and the offset $\rho(i,i)-\rho(j,j)$ for the two random initial condition realizations at $\eta=10$. As expected, the offset is comparable to the coherence in the two random initial condition, similarly to the analytical results obtained for $Q=1$. 
}
\end{center}
\end{figure}

In Fig.~\ref{fig:coherences_steadystate} we show that the steady state is reached for the two random initial conditions by comparing the evolutions of the strongest coherent term and a diagonal offset.

To conclude, we provide evidence that the quantum diffusion behaviour observed for $Q=1$ is indeed general even for $Q\neq1$. The diffusive behaviour is implied from the evolution equations for $Q=2$.

\end{document}